\documentclass[acmtog]{acmart}

\usepackage{booktabs} 

\usepackage[ruled]{algorithm2e} 

\SetAlFnt{\small}
\SetAlCapFnt{\small}
\SetAlCapNameFnt{\small}
\SetAlCapHSkip{0pt}
\IncMargin{-\parindent}

\setcopyright{rightsretained}

\usepackage{wrapfig}
\usepackage{floatflt}
\usepackage{amsmath}
\usepackage{mathrsfs}

\DeclareMathOperator*{\argmin}{arg\,min}

\def\RR{\mathbb{R}}

\newcommand{\figspace}{\vspace{0.0cm}}

\newcommand{\secspace}{\vspace{-0.00cm}}
\newcommand{\hide}[1]{}
\newcommand{\SIGASIA}[1]{}

\newcommand{\BF}[1]{\mathbf{#1}}

\newcommand{\PP}[2]{\frac{\partial #1}{\partial #2}}
\newcommand{\PPI}[2]{\partial #1 / \partial #2}
\newcommand{\PPP}[2]{\frac{\partial^2 #1}{\partial #2^2}}
\newcommand{\PPPI}[2]{\partial^2 #1 / \partial #2^2}
\newcommand{\PPPP}[3]{\frac{\partial^2 #1}{\partial #2 \partial #3}}
\newcommand{\PPPPI}[3]{\partial^2 #1 / (\partial #2 \partial #3)}

\begin{document}


\title{Modeling of Personalized Anatomy using Plastic Strains} 

\author{Bohan Wang}
\affiliation{%
  \institution{University of Southern California}
  \city{Los Angeles}
  \state{CA}
  \country{USA}
}
\email{bohanwan@usc.edu}

\author{George Matcuk}
\affiliation{%
  \institution{University of Southern California}
  \city{Los Angeles}
  \state{CA}
  \country{USA}
}
\email{matcuk@usc.edu}

\author{Jernej Barbi\v{c}}
\affiliation{%
  \institution{University of Southern California}
  \city{Los Angeles}
  \state{CA}
  \country{USA}
}
\email{jnb@usc.edu}

\begin{abstract}
We give a method for modeling solid objects undergoing large spatially 
varying and/or anisotropic strains, 
and use it to reconstruct human anatomy from medical images.
Our novel shape deformation method uses plastic strains
and the Finite Element Method to successfully model shapes undergoing 
large and/or anisotropic strains, specified by sparse point constraints 
on the boundary of the object.
We extensively compare our method to standard second-order shape
deformation methods, variational methods and surface-based methods
and demonstrate that our method avoids the spikiness, wiggliness
and other artefacts of previous methods.
We demonstrate how to perform such shape deformation both for attached
and un-attached (``free flying'') objects, using a novel method to
solve linear systems with singular matrices with a known nullspace.
While our method is applicable to general large-strain shape deformation
modeling, we use it to create personalized 3D triangle and volumetric meshes
of human organs, based on MRI or CT scans.
Given a medically accurate anatomy template of a generic individual, we 
optimize the geometry of the organ to match the MRI or CT scan 
of a specific individual. Our examples include human hand muscles, 
a liver, a hip bone, and a gluteus medius muscle (``hip abductor''). 
\end{abstract}



\begin{teaserfigure}
\centerline{\includegraphics[width=1.0\hsize]{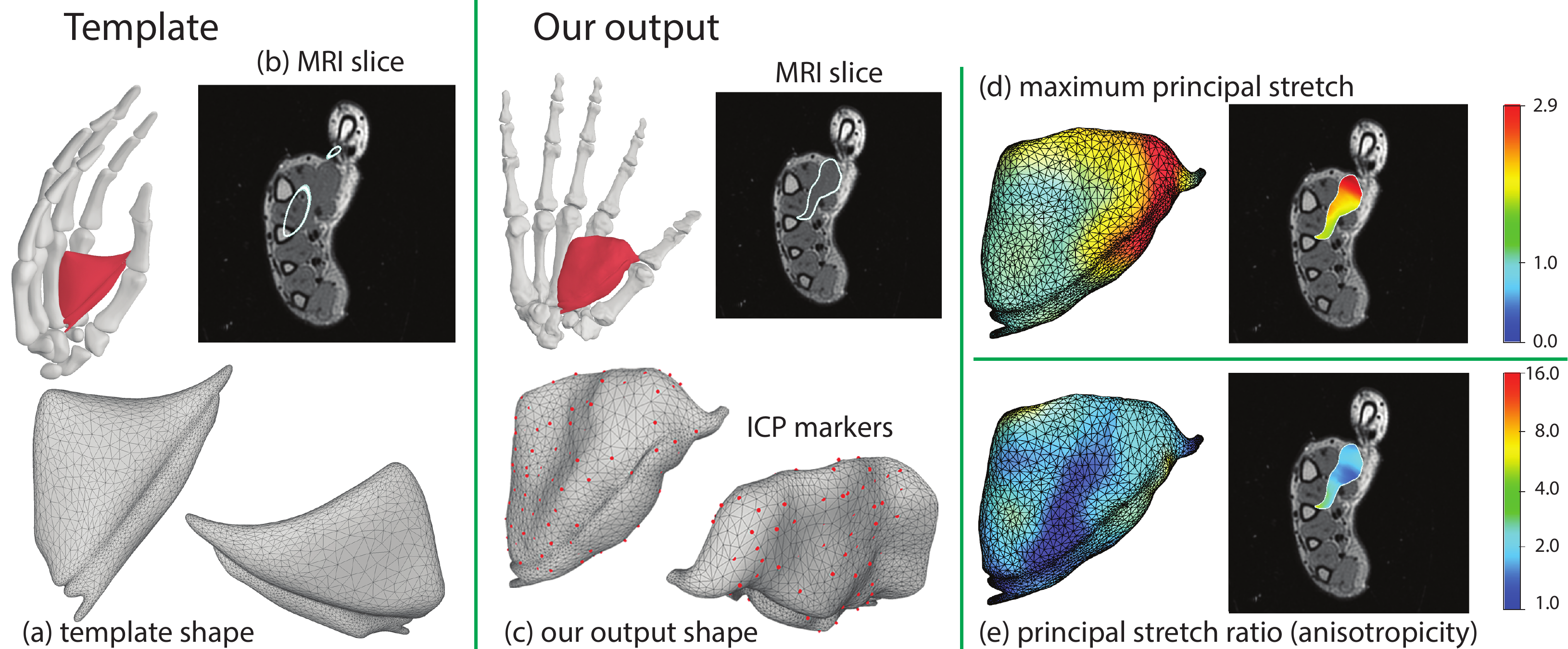}}
\vspace{-0.1cm}
  \caption{\textbf{We optimized the shape of this hand muscle (Adductor Pollicis) to match the MRI scan.}
(a) template shape~\cite{Dundee};
(b) representative MRI slice~\cite{Wang:2019:HMA}; we rigidly aligned the template mesh
onto the markers in the MRI scan, producing the white contour which is obviously incorrect 
(deeply penetrates the bone; and extends out of the volume of the MRI-scanned hand);
(c) our output shape, optimized to the MRI scan; the white contour now matches the scan;
(d,e) large anisotropic spatially varying strains accommodated by our method 
(d: maximum principal stretch; e: ratio between maximum and minimum principal stretch).
}
\label{fig:teaser}
\end{teaserfigure}

\maketitle

\secspace
\section{Introduction}
\secspace

Modeling and simulating human anatomy is 
very important in many applications in computer graphics, animation, medicine, film and
real-time systems such as games and virtual reality.
In this paper, we demonstrate how to model anatomically realistic 
\emph{personalized} three-dimensional shapes of human organs, 
based on medical images of a real person, such as Magnetic Resonance Imaging (MRI) or Computed Tomography (CT).
Such modeling is crucially important for personalized medicine. For example, after scanning the patient with an MRI or CT scanner, 
doctors can use the resulting 3D meshes to perform pre-operative surgery planning. 
Such models are also a starting point for anatomically based  human simulation 
for applications in computer graphics, animation and virtual reality.
Constructing volumetric meshes that match an organ in a medical image 
can also help with building volumetric correspondences between multiple MRI or CT scans of the same 
person~\cite{Rhee:2011:SBV}, e.g., for medical education purposes.

Although the types, number and function of organs in the human body are generally the same for any human, the shape of 
each individual organ varies greatly from person to person, due to the natural variation across the human population. 
The shape variation is substantial: any two individuals' 
organs $\Omega_1\subset\RR^3$ and $\Omega_2\subset\RR^3$ generally vary by a non-trivial 
shape deformation function $\Phi : \Omega_1 \to \Omega_2$ that often consists of large and spatially varying anisotropic strains
(see Figure~\ref{fig:teaser}). 
By ``large anisotropic strain'', we mean that the singular values of the 
3x3 gradient matrix of $\Phi$ are both different to each other and substantially different from 1.0, i.e.,
the material locally stretches (or compresses) by large amounts; and this amount is different in different directions and 
varies spatially across the model. 

We tackle the problem of how to model such large shape variations, 
using volumetric 3D medical imaging (such as MRI or CT scan),
and a new shape deformation
method capable of modeling large spatially varying anisotropic strains. 
We note that the boundary between the different organs in  medical images 
is often blurry. For example, in an MRI of a human hand, the muscles often ``blend'' into each other and
into fat without clear boundaries; a CT scan has even less contrast. We therefore manually select as many reliable points ("markers") as possible on the boundary of the organ in the medical image; some with correspondence (``landmark constraints'') to the same anatomical landmark in the template organ, and 
some without (``ICP constraints''). Given a template volumetric mesh of an organ of a generic individual,
a medical image of the same organ of a new individual, 
and a set of landmark and ICP (Iterative Closest Point) constraints, our paper asks how to deform the template mesh
to match the medical image. 

\begin{figure}[!t]
\centering
\includegraphics[width=1.0\hsize]{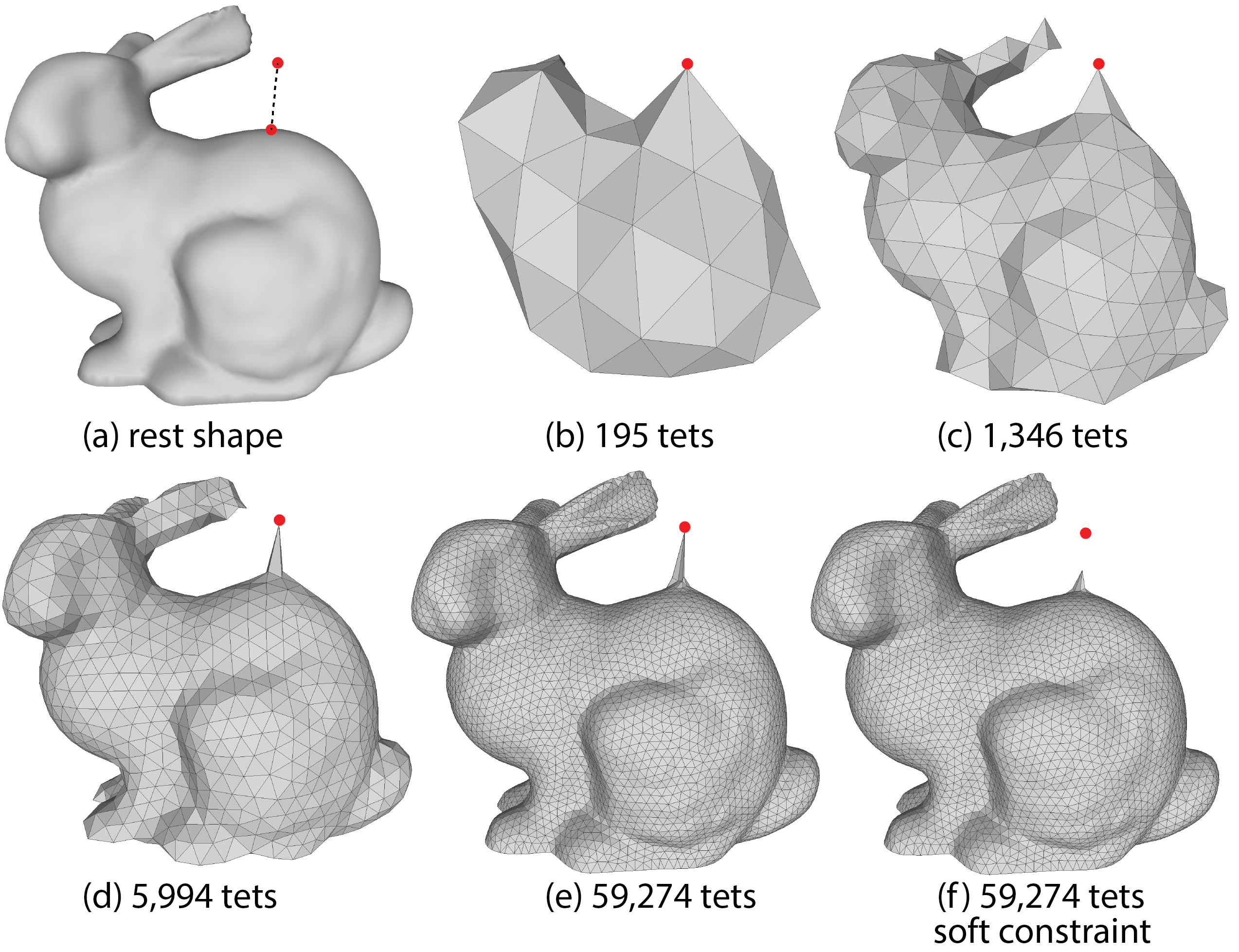}
\figspace{}
\vspace{-0.6cm}
\caption{ {\bf Second-order methods produce spiky outputs as the tet mesh is refined.}
Here, we show the output of an FEM static solver under a single hard point constraint
(seen in (a)), i.e., we minimize the FEM elastic energy under the shown hard point constraint.
Bunny is fixed at the bottom.
Poissons's ratio is 0.49. As we increase the tet mesh resolution in (b)-(e), the spike becomes
progressively narrower, which is undesirable. Changing the elastic 
stiffness (Young's modulus) 
of the bunny changes nothing. Converting the constraint into a soft 
spring constraint also does not help (f); now, the constraint is not even satisfied.
}
\vspace{-0.25cm}
\label{fig:FEM-spikes}
\end{figure}
Our first attempt to solve this problem was to use standard shape 
deformation methods commonly used in computer graphics, such as
as-rigid-as-possible energy (ARAP)~\cite{Sorkine:2007:ARA},
bounded biharmonic weights (BBW)~\cite{Jacobson:2011:BBW},
biharmonic weights with linear precision (LBW)~\cite{Wang:2015:LSD},
and a finite element method static solver (FEM)~\cite{Barbic:2009:DOA}.
As we show in Section~\ref{sec:standardShapeDeformation}, 
none of these methods was able to capture the large strains
observed in medical images. Namely, these standard methods 
either cannot model point constraints when the shape undergoes large spatially varying strains,
or introduce excessive curvature. 
For example, in the limit where the tetrahedral mesh is refined to finer and finer tets, 
the FEM static solver produces a spike output (Figure~\ref{fig:FEM-spikes});
and similar limitations apply to the other methods.

Instead, we give a new shape deformation method that uses plastic strains
and the Finite Element Method to successfully model shapes undergoing
large and/or anisotropic strains, controlled by the sparse landmark and ICP point 
constraints on the boundary of the object.
In order to do so, we formulate a nonlinear optimization problem
for the unknown \emph{plastic deformation gradients} 
of the template shape $\Omega_1,$ such that under these gradients, the
shape $\Omega_1$ transforms into a shape that matches the medical image landmark
and ICP constraints. The ICP constraints are handled by properly
incorporating the ICP algorithm into our
method. We note that in solid mechanics, plastic deformation
gradients are a natural tool to model large volumetric shape variations. We are, however, unaware
of any prior work that has used plastic deformation gradients and the
Finite Element Method to model large-strain shape deformation. 

In order to make our method work, we needed to overcome several 
numerical obstacles. The  large-strain shape optimization problem 
is highly nonlinear and cannot be reliably solved with off-the-shelf 
optimizers such as the interior point method~\cite{Knitro}. 
Furthermore, a naive
solution requires solving large dense linear systems of equations.
We demonstrate how to adapt the Gauss-Newton optimization 
method to robustly and efficiently solve our shape deformation problem, and how
to numerically avoid dense linear systems.
In order to optimize our shapes, we needed to derive analytical 
gradients of internal elastic forces 
and the tangent stiffness matrix with respect to the plastic strain, 
which will be useful for further work on using plasticity for 
optimization and design of 3D objects.
In addition, we address objects that are attached to other objects, 
such as a hand muscle
attached to one or more bones; as well as un-attached objects. 
An example of an un-attached object is a liver, 
where the attachments to the surrounding tissue certainly exist, 
but are not easy to define.  It is practically easier to just model 
the liver as an un-attached object.
In order to address un-attached objects, we give a novel numerical method to
solve linear systems with singular matrices with a known nullspace.
Such linear systems are commonly encountered in applications in
geometric shape modeling and nonlinear elastic simulation.
Our examples include human hand muscles, a liver,
a hip bone and a hip abductor muscle (``gluteus medius''),
all of which undergo substantial and non-trivial shape
change between the template and the medical image.

\secspace
\section{Related Work}
\secspace

In this section, we introduce closely related work 
and discuss the relationship to our work.

\subsubsection*{Geometric shape modeling}

Geometric shape modeling is an important topic in computer graphics research;
e.g., see the Botsch and Sorkine~\cite{Botsch:2008:OLV} survey and
the SIGGRAPH course notes by Alexa et al.~\cite{Alexa:2006:ISM}.
Popular methods include 
variational methods~\cite{Botsch:2004:AIF},
Laplacian surface editing~\cite{Sorkine:2004:LSE}, 
as-rigid-as-possible (ARAP) 
deformation~\cite{Igarashi:2005:ARA,Sorkine:2007:ARA}, 
coupled prisms~\cite{Botsch:2006:PCP}
and partition-of-unity methods such 
as bounded biharmonic weights (BBW)~\cite{Jacobson:2011:BBW} and
biharmonic weights with linear precision~\cite{Wang:2015:LSD};
we provide a comparison 
in Section~\ref{sec:standardShapeDeformation} and in several
other Figures in the paper.
Our method reconstructs the surface shape from a set of 
un-oriented point observations; this goal is similar 
to variational implicit surface 
methods~\cite{Turk:1999:STU,Huang:2019:VIP}; we give a comparison
in Section~\ref{sec:results}.
Point clouds can also be used to optimize 
rest shapes~\cite{Twigg:2011:OSS} and material properties
of 3D solids~\cite{Wang:2015:DCM}. Such a method cannot be applied
to our problem because it assumes a 4D dense point cloud input;
whereas we assume 3D sparse point inputs as commonly encountered
in medical imaging.
Point constraint artifacts of second-order methods can be addressed using 
spatial averaging~\cite{Bergou:2007:TTD,Kavan:2011:PIU};
however this requires specifying the averaging functions
(often by hand) and, by the nature of averaging, causes the constraints to
be met only approximately. Our method can meet the constraints
very closely (under $0.5$ mm error in our examples), 
i.e., in the precision range of the medical scanners.

\subsubsection*{Plasticity}

Elastoplastic simulations are widely used in computer
animation. O'Brien et al.~\shortcite{Obrien:2002:GMA} and
Muller and Gross~\shortcite{Mueller:2004:IVM} used an additive
plasticity formulation, whereas Irving et al.~\cite{Irving:2004:IFE}
presented a multiplicative formulation and argued that it is
better for handling large plasticity;
we adopt multiplicative formulation in our work.
The multiplicative model was used in many subsequent publications
to simulate plasticity, 
e.g.,~\cite{Bargteil:2007:FEM,Stomakhin:2013:AMP,Chen:2018:PBF}.
Because plasticity models shapes that undergo permanent and large deformation, it is in principle a natural choice also for  geometric shape modeling. However, such an application  is not straightforward: an
incorrect choice of the optimization energy will produce degenerate outputs,   elastoplastic simulations in equilibrium  lead to linear systems with singular matrices, optimization requires second-order derivatives for fast convergence, and easily produces large linear systems with dense matrices. We present a solution to these obstacles. To the best of our knowledge,  we are the first paper to present such a comprehensive approach for using plasticity for geometric shape modeling with large and anisotropic strains.

\subsubsection*{Anatomically based simulation}

Anatomically based simulation of the human body has been explored 
in multiple publications. For example, researchers
simulated human facial muscles~\cite{Sifakis:2005:ADF},
the entire upper human body~\cite{Lee:2009:CBM},
volumetric muscles for motion control~\cite{Lee:2018:DMA}
and hand bones and soft tissue~\cite{Wang:2019:HMA}.
Anatomically based simulation is also popular in 
film industry~\cite{Tissue}.
Existing papers largely simulate generic humans because it is not 
easy to create accurate anatomy personalized to each specific person. 
Our method can provide such an input anatomy, 
based on a medical image of any specific new individual. 

\subsubsection*{Medical image registration}

Deformable models are widely used in medical image analysis~\cite{McInerney:2008:DM}.
Extracting quality anatomy geometry from medical images is difficult. 
For example, Sifakis and Fedkiw~\shortcite{Sifakis:2005:ADF} 
reported that it took them ``six months'' (including implementing the tools) 
to extract the facial muscles from the visible human dataset~\cite{VisibleHuman}, 
and even with the tools implemented it would still take ``two weeks''.
With our tools, we are able to extract all the 17 muscles of the human
hand in 1 day (including computer and user-interaction time).
Bones generally have good contrast against the surrounding tissue
and can be segmented using active contour methods~\cite{Szekely:1996:SO2}
or Laplacian-based segmentation~\cite{Grady:2006:RWF,Wang:2019:HMA}.
For bones, it is therefore generally possible to obtain a ``dense'' set
of boundary points in the medical image.
Gilles et al.~\cite{Gilles:2010:CAA} used this to deform 
template skeleton models to match a subject-specific MRI scan and posture.
They used the ARAP energy and deformed surface meshes.
In contrast, we give a method that is suitable for soft tissues where the image contrast is 
often low (our hand muscles and liver examples) and that accommodates
volumetric meshes and large volumetric scaling variations between the template and the subject. 
If one assumes that the template mesh comes with a registered MRI scan 
(or if one manually creates a template mesh that matches a MRI scan), musculoskeletal reshaping 
becomes more defined because one can now use the pair of MRI images, namely the template and target,
to aid with reshaping the template mesh~\cite{Gilles:2006:AMO,Schmid:2009:MSM,Gilles:2010:MMS}. 
The examples in these papers demonstrate non-trivial musculoskeletal reshaping involving translation and spatially varying
large rotations with a limited amount of volumetric stretching (Figure 14 in~\cite{Gilles:2010:MMS}).
This is consistent with their choice of the similarity metric between the template and output shapes:
their reshaping energy tries to keep the distance of the output mesh to the medial axis the same 
as the distance in the template~\cite{Gilles:2010:MMS}, which biases the output against volume growth. 
For bones, a similar idea was also presented in~\cite{Schmid:2008:MBS,Schmid:2011:AGF}, 
where they did not use a medial-axis
term to establish similarity to a source mesh, but instead relied on a PCA prior on the shapes of bones, 
based on a database of 29 hip and femur bone shapes. 
Our work does not require any pre-existing database of shapes.
Because our method uses plasticity, it can accommodate large and spatially varying volumetric stretching between
the template and the subject. We do not need a medical image for the template mesh. We 
only assume that the template mesh is plausible. Of course, the template mesh itself might have been
derived from or inspired by some MRI or CT scan, but there is no requirement that it matches any such scan.

\subsubsection*{Anatomy Transfer}

Recently, great progress has been made on anatomically based
simulations of humans. Anatomy transfer has
been pioneered by Dicko and colleagues~\cite{Dicko:2013:AT}.
Anatomical muscle growth and shrinkage
have been demonstrated in the ``computational bodybuilding'' 
work~\cite{Saito:2015:CBA}.
Kadlecek et al.~\shortcite{Kadlecek:2016:RPA} demonstrated 
how to transfer simulation-ready anatomy to a novel human, 
and Ichim et al.~\cite{Ichim:2017:PPF} gave a 
state-of-the-art pipeline for anatomical simulation of human faces.
Anatomy transfer and a modeling method such as ours are complementary because the former provides the means to interpolate known anatomies to new subjects, whereas the latter provides a means to create the anatomies in the first place. Namely, anatomy transfer requires a quality anatomy template to serve as the source of anatomy transfer, which brings up the question of how one obtains such a template. Human anatomy is both extremely  complex for each specific subject, and exhibits
large variability in geometry across the population. Accurate templates can therefore only be created by matching them to medical images. Even if one creates such a template, new templates will always be needed to model the anatomical variability across the entire population; and this requires an anatomy modeling method such as ours.  

\secspace
\section{Shape Deformation with Large Spatially Varying Strains}
\label{sec:shapeDeformation}
\secspace

\begin{figure}[!t]
\centering
\includegraphics[width=1.0\hsize]{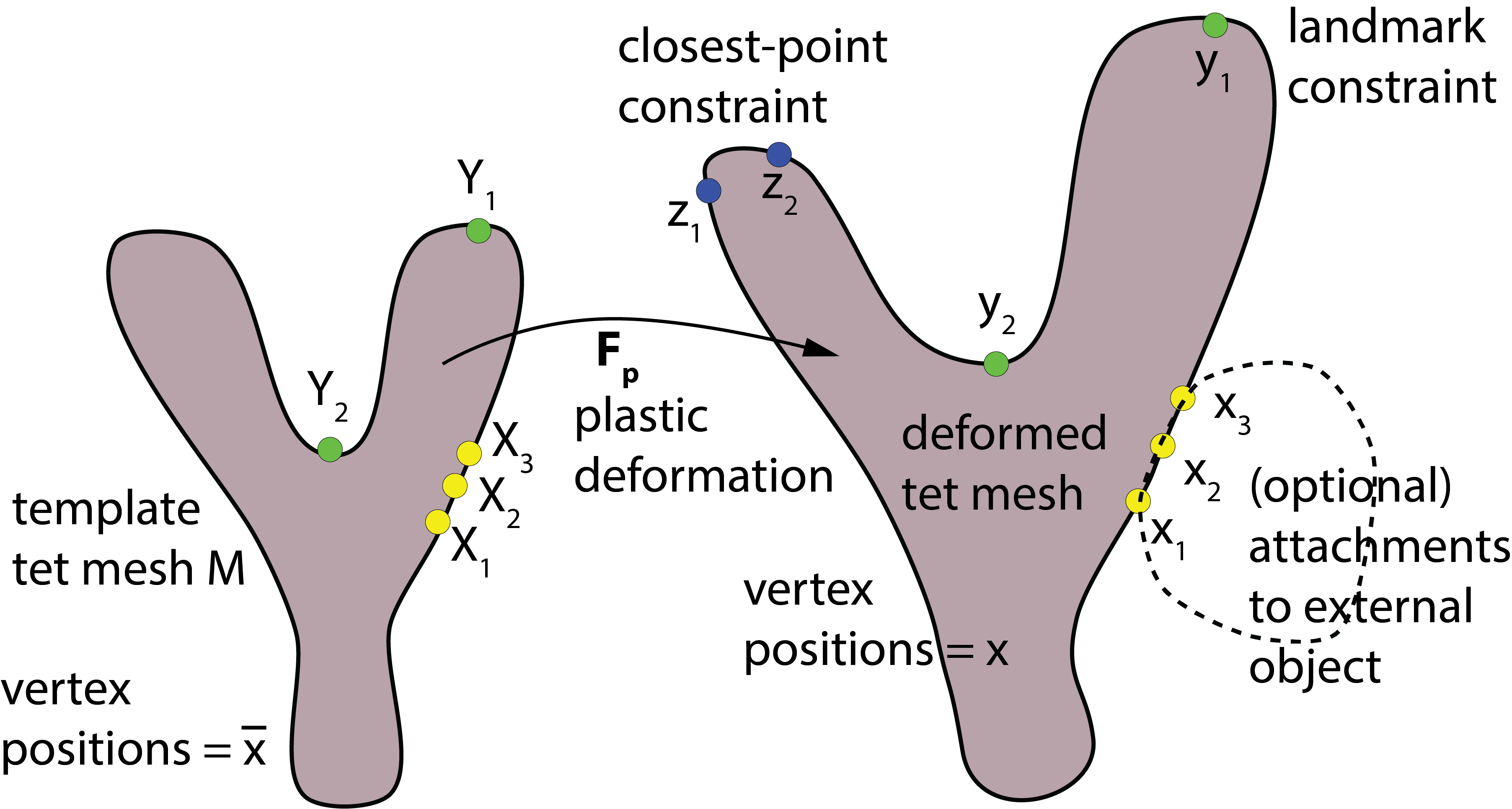}
\figspace{}
\vspace{-0.6cm}
\caption{ {\bf Our shape deformation setup. }
Our optimization discovers plastic strains $F_p$
and vertex positions $x$ so that the model is in 
elastic equilibrium under the attachments, while meeting the medical image
landmark and closest point constraints as closely as possible. The presence of attachments is optional; our work handles both attached and un-attached objects. 
}
\vspace{-0.25cm}
\label{fig:shapeDeformation}
\end{figure}

Given a template tet mesh of a soft tissue organ for a generic individual, 
as well as known optional attachments of the organ to other objects, 
our goal is to deform the tet mesh to match a medical image of the organ of a new individual. 
We use the term ``medical image'' everywhere in this paper because this is standard terminology; 
this does not refer to an actual 2D image, but to the 3D medical volume.
We now describe how we mathematically model the attachments and medical image constraints.

\subsection{Attachments and medical image constraints}
\label{sec:attachments}

We start with a template organ tet mesh $\mathcal{M}.$ We denote 
its vertex positions by $\BF{\bar{x}}\in\RR^{3n},$ 
where $n$ is the number of tet mesh vertices. 
In this paper, we use \textbf{bold} text to represent the quantities 
for the entire mesh and non-bold text to represent the quantities 
for a single vertex or element in the FEM mesh.
We would like to discover vertex positions $\BF{x}\in\RR^{3n}$ such that the
organ shape obeys the  attachments to other 
organs (if they exist), and of course the medical image. 
The  attachments are modeled by known
material positions $X_i\in\mathcal{M},\,i=1,\ldots,t$ that have
to be positioned at known 
world-coordinate positions $x_i\in\RR^3,i=1,\ldots t$
(Figure~\ref{fig:shapeDeformation}).
The medical image constraints come in two flavors.
First, there are \emph{landmark constraints} 
whereby a point on a template organ
is manually corresponded to a point in the medical image, based
on anatomy knowledge. Namely, landmark constraints are
modeled as material positions $Y_i\in\mathcal{M},\,i=1,\ldots,q,$
that are located at known world-coordinate positions 
$y_i\in\RR^3$ in the medical image. Observe that landmark constraints
are mathematically similar to attachments. However, they
have a different physical origin: attachments are a physical
constraint that is pulling the real-world organ to a known 
location on another (fixed, un-optimized) object; for example, a muscle is attached to a bone.
With landmarks, there is no such physical 
force in the real-world; namely, landmarks (and also closest-point
constraints) are just medical image observations.  

The second type of medical image 
constraints are 
\emph{closest-point constraints} (``ICP markers'').
They are given by known world-coordinate 
positions $z_i\in\RR^3,i=1,\ldots r$
that have to lie on the surface of  the deformed tet mesh.
Locations $z_i$ are easier to select in the medical image 
than the landmarks because there is no need
to give any correspondence. As such, they require
little or no medical knowledge, and can be easily selected in large numbers. 
We simply went through the medical image slices and 
selected clear representative points on the organ boundary. We then visually compared the template and the target shape inferred by the
ICP marker cloud. This guided our positioning of the landmarks, which we place on anatomically
``equal'' positions in the template and the medical image. We consulted a medical doctor to help us interpret medical images, such as identifying muscles in the scan, clarify ambiguous muscle boundaries, placing difficult markers and attachments, and disambiguating tendons.

\subsection{Plastic deformation gradients}
\label{sec:plasticDeformationGradients}

We model shape deformation using plastic deformation gradients,
combined with a (small) amount of elastic deformation. 
In solid mechanics, plasticity is the tool to model large 
shape variations of objects, making it very suitable to model 
our desired shape deformation with large strains. Unlike
using the elastic energy directly (without plasticity), plastic
deformations have the advantage that they can arbitrarily and 
spatially non-uniformly and anisotropically
scale the object. There is also no mathematical requirement
that they need to respect volume preservation constraints. 
This makes plastic deformations a powerful tool to model shapes. 
Our key idea is to find a plastic deformation
gradient $F_p$ at each tet of $\mathcal{M},$ such that the 
FEM equilibrium shape under $\BF{F_p}$ and any attachments
matches the medical image observations.
Figure~\ref{fig:shapeDeformation} 
illustrates our shape deformation setting. In order to do so, we need to discuss the elastic energy and forces in the presence of plastic deformations, which we do next.

\begin{figure*}[!ht]
        \centering
        \includegraphics[width=1.0\hsize]{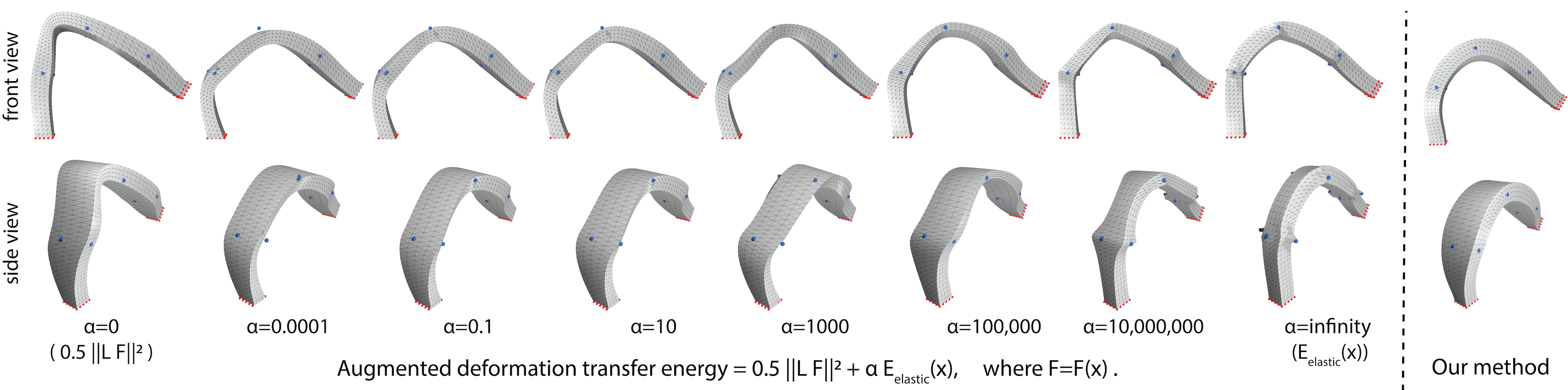}
        \figspace{}
        \vspace{-0.75cm}
        \caption{ {\bf Comparison to augmented deformation transfer.} The beam's attachments (red) 
cause the beam to bend, whereas the ICP markers (blue) cause it to stretch 2x in one of the two transverse directions.
Our method can easily recover such a shape deformation, 
whereas deformation transfer~\cite{Sumner:2004:DTF} cannot, even if augmented with an elastic energy.} 
        \vspace{-0.25cm}
        \label{fig:deformationTransfer}
\end{figure*}

\begin{figure}[!t]
\centering
\includegraphics[width=1.0\hsize]{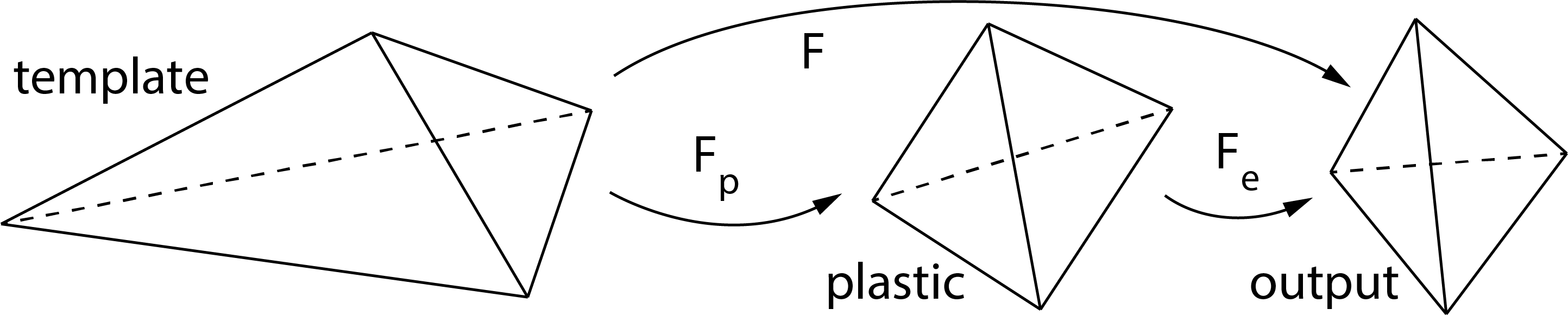}
\figspace{}
\vspace{-0.6cm}
\caption{ {\bf Plastic and elastic deformation gradient for a single tet.}
}
\vspace{-0.25cm}
\label{fig:FpAndFe}
\end{figure}
Plastic strain is given by a $3\times 3$ matrix 
$F_p$ at each tetrahedron of $\mathcal{M}.$
For each specific deformed shape $\BF{x}\in\RR^{3n},$ 
one can define and compute the deformation gradient $F$ between 
$\BF{\overline{x}}$ and $\BF{x}$ at each tet~\cite{Mueller:2004:IVM}.
The elastic deformation gradient $F_e$ can then be defined 
as $F_e=F F_p^{-1}$~\cite{Bargteil:2007:FEM} (see Figure~\ref{fig:FpAndFe}).
Observe that for any shape $\BF{x},$ there exists a corresponding
plastic deformation gradient $\BF{F_p}$ such that $\BF{x}$ is the elastic
equilibrium under $\BF{F_p};$ namely $\BF{F_p}=\BF{F}.$ This means that
the space of all plastic deformation 
gradients $\BF{F_p}$ is expressive enough to capture all shapes $\BF{x}.$
The elastic energy of a single tet is defined as 
\begin{equation}
\mathscr{E}(F_p,x)=V(F_p) \psi(x, F_p)=V(F_p) 
\psi\bigl(F(x) F_p^{-1}\bigr),
\end{equation}
where $V$ is the rest volume of the tet under the 
plastic deformation $F_p,$ and 
$\psi$ is the elastic energy density function.
We have $V = |F_p|\, V_0,$ 
where $V_0$ is the tet's volume in $\mathcal{M}$, 
and $|F_p|$ is the determinant of the matrix $F_p.$
Elastic forces equal 
$\BF{f_\textrm{e}}(\BF{F_p}, \BF{x}) = 
d \mathscr{E}(\BF{F_p}, \BF{x}) / d \BF{x}.$
When solving our optimization problem to compute $\BF{F_p}$
in Section~\ref{sec:solveShapeOptimization},
we will need the first and second derivatives 
of $\mathscr{E}(F_p,x)$ with respect to $x$ and $F_p.$ 
We provide a complete derivation of these terms in 
Appendix~\ref{app:appendixDerivatives}.

\begin{figure*}[!t]
\centering
\includegraphics[width=1.0\hsize]{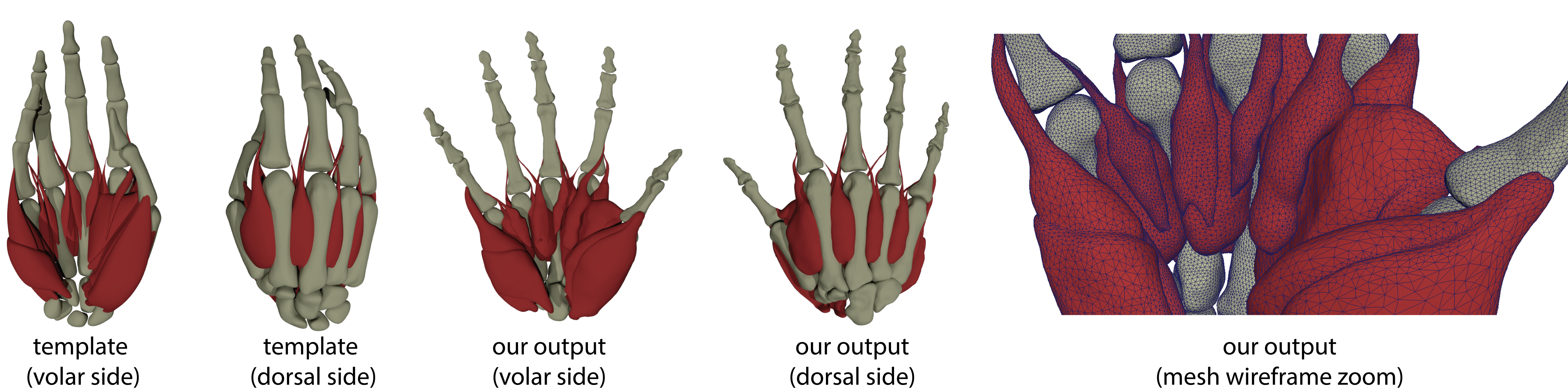}
\figspace{}
\vspace{-0.6cm}
\caption{ {\bf 17 muscles of the human hand extracted from MRI.}
Observe that the template hand is bigger than the scanned hand.
Pose is also different. 
Our method solves this using bone attachments. 
}
\vspace{-0.25cm}
\label{fig:musclesInMRI}
\end{figure*}

Our method supports any isotropic hyperelastic energy density function $\psi.$
In our examples, we use the isotropic stable neo-Hookean elastic 
energy~\cite{Smith:2018:SNF},
because we found it to be stable and sufficient for our examples.
Note that we do model anisotropic plastic strains (and this is crucial for our method), 
so that our models can stretch by different amounts in different directions.
Observe that plastic strains 
are only determined up to a rotation. Namely, let $F_p$ be a plastic strain
(we assume $\textrm{det}(F_p)>0;$ i.e., no mesh inversions),
and $F_p=QS$ be the polar decomposition where $Q$ is a rotation and $S$
a $3\times 3$ \emph{symmetric} matrix. Then, $F_p$ and $S$ are the ``same'' plastic
strain: the resulting elastic deformation gradients differ only by a rotation, and hence,
due to isotropy of $\psi,$ produce the same elastic energy and elastic forces. 
Note that it is not required that rotations $Q$ match in any way at adjacent tets.
We do not need to even guarantee that $F_p$ globally 
correspond to any specific ``rest shape'',
i.e., the $F_p$ are independent of each other and may be inconsistent. 
This gives plastic deformation gradient modeling a lot of flexibility.
Hence, it is sufficient to model plastic strains as symmetric $3\times 3$
matrices. We can therefore model $F_p$ as a symmetric matrix and 
parameterize it using a vector $s \in \RR^6,$
\begin{equation}
F_p= 
\begin{bmatrix}
s_1 & s_2 & s_3 \\
s_2 & s_4 & s_5 \\
s_3 & s_5 & s_6
\end{bmatrix}.
\label{eq:Fp-s}
\end{equation}
We model plasticity globally using 
a vector $\BF{s}\in\RR^{6m},$ where $m$ is the number of tets in $\mathcal{M}.$
We note that Ichim et al.~\shortcite{Ichim:2017:PPF} used such a 6-dimensional
parameterization to model facial muscle activations.
In our work, we use it for general large-strain shape modeling. 
Our application and optimization energies
are different, e.g, Ichim et al.~\shortcite{Ichim:2017:PPF} 
causes muscle shapes to follow a prescribed muscle firing field, and biases
principal stretches to be close to 1. Furthermore, we address the singularities
arising with un-attached objects.

\subsection{Shape deformation of attached objects}

We now formulate our shape deformation problem. 
We first do so for attached objects. 
An object is ``attached'' if there are sufficient attachment forces to remove all six rigid degrees of freedom, which is generally satisfied if there are at least three attached non-colinear vertices. We find the organ's shape that matches the attachment 
and the medical image constraints by finding a plastic 
strain $F_p$ at each tet, as well as static equilibrium 
tet mesh vertex positions $\BF{x}$ under the attachments
and plastic strain $\BF{F_p},$ so that the medical image 
observations are met as closely as possible,
\begin{gather}
\label{eq:objectiveFunction}
\argmin_{\BF{s},\,\BF{x}}\quad ||\BF{L}\, \BF{s}||^2 + 
\alpha \mathcal{E}_\textrm{MI}(\BF{x}) +
\beta\mathcal{E}_\textrm{a}(\BF{x}) ,\\
\textrm{subject\ to:\quad }
\BF{f_\textrm{e}}(\BF{F_p}(\BF{s}), \BF{x}) + 
\BF{f_\textrm{a}}(\BF{x})  = 0,
\label{eq:constraints}
\end{gather}
where $\alpha\geq 0$ and $\beta\geq 0$
are scalar trade-off weights,
and $\BF{L}$ is the \emph{plastic strain Laplacian}.
We define $\BF{L}$ as essentially the tet-mesh Laplacian operator on the tets,
6-expanded to be able to operate on entries of $s$ at each tet 
(precise definition is in Appendix~\ref{app:laplacian}). 
The Laplacian term enforces the smoothness of $\BF{F_p},$
i.e., $F_p$ in adjacent tets should be similar to each other.
The second equation enforces the elastic equilibrium of the model 
under plastic strains $\BF{F_p}$ and under the attachment forces $\BF{f_a}.$
As such, our output shapes are always in static equilibrium
under the plastic strains $\BF{F_p},$ and both this equilibrium shape $\BF{x}$
and $\BF{F_p}$ are optimized together; this is the key aspect of our work.
The first equation contains the smoothness 
and the medical image (MI) observations;
we discuss the attachment energy $\mathcal{E}_\textrm{a}$ 
in the next paragraph.
The medical image energy measures how closely $\BF{x}$ matches the medical 
image constraints,
\begin{equation}
\mathcal{E}_\textrm{MI}(\BF{x}) = 
\sum_{i=1}^q ||S \BF{x} - y_i||^2 +  
\sum_{i=1}^r ||z_i - \textrm{closestPoint}(\BF{x}, z_i)||^2,
\end{equation}
where $S$ is the interpolation matrix that selects
$Y_i,$ namely $S \bar{x} = Y.$
The function $\textrm{closestPoint}(\BF{x}, z_i)\in\RR^3$
computes the closest point to $z_i\in\RR^3$ on the surface
of the tet mesh with vertex positions $\BF{x}.$

Our treatment of attachments in Equations~\ref{eq:objectiveFunction}
and~\ref{eq:constraints} deserves a special notice. 
Equation~\ref{eq:constraints} is consistent with
our setup: we are trying to explain the medical images by
saying that the organ has undergone a plastic deformation due to
the variation between the template and captured individual. 
The shape observed in the medical image is due to this plastic 
deformation and the  attachments. We formulate attachment 
forces in Equation~\ref{eq:constraints}
as a ``soft'' constraint, i.e., $\BF{f_{a}}(\BF{x})$
is modeled as (relatively stiff) springs pulling the attached organ points
to their position on the external object. This soft constraint could in principle be replaced
for a hard constraint where the attached positions 
are enforced exactly. We use soft constraints in our examples
because they provide additional control to balance attachments against medical image landmarks and ICP markers. 
These inputs are always somewhat inconsistent because it is impossible 
to place them at perfectly correct anatomical locations, 
due to medical imaging errors. Hence, it is useful to have some leeway in adjusting 
the trade-off between satisfying each constraint type. 
With soft constraints, it is important to keep the spring coefficient in $\BF{f_a}(\BF{x})$ high
so that constraints are met very closely (under $0.5$ mm error in our examples). 

As per the attachment energy $\mathcal{E}_\textrm{a},$
we initially tried solving the optimization problem of 
Equations~\ref{eq:objectiveFunction} and~\ref{eq:constraints} without it.
This seems natural, but actually did not work. 
Namely, without $\mathcal{E}_\textrm{a},$
there is nothing in 
Equations~\ref{eq:objectiveFunction} and~\ref{eq:constraints}
that forces the plastic strains to  reasonable values. 
The optimizer is free to set $\BF{F_p}$ to arbitrarily extreme 
values, and then find a static equilibrium $\BF{x}$ under the attachment forces.
In our outputs, we would see smooth nearly tet-collapsing plastic strains that
result in a static equilibrium $\BF{x}$ whereby the medical image constraints were
nearly perfectly satisfied. Obviously, this is not a desired outcome.
Our first idea was to add a term that penalizes the elastic
energy $\mathscr{E}(\BF{F_p},\BF{x})$ to Equation~\ref{eq:objectiveFunction}.
Although this  worked in simple cases, it makes the expression
in Equation~\ref{eq:objectiveFunction} generally nonlinear.
Instead, we opted for a simpler and more easily  computable
alternative, namely add the elastic spring energy of 
all attachments, $\mathcal{E}_\textrm{a}.$ This keeps
the expression in Equation~\ref{eq:objectiveFunction} quadratic
in $\BF{x}$ and $\BF{F_p},$ which we exploit in 
Section~\ref{sec:solveShapeOptimization} for speed.
Observe
that $\mathcal{E}_\textrm{a}$ behaves similarly to the
elastic energy: if the plastic strain causes a rest shape that is far
from the attachment targets, then both $\mathcal{E}_\textrm{a}$
and the elastic energy will need to ``work'' to bring the shape $x$
to its target attachments.
Similarly, if the plastic strain already did most of the work 
and brought the organ close to its target, then neither $\mathcal{E}_\textrm{a}$ nor 
the elastic energy will need to activate much.

Because our units are meters and we aim to satisfy constraints closely,
we typically use weights close to $\alpha=10^9$ and $\beta=10^8$ in our examples.
The weights $\alpha$ and $\beta$ permit adjusting the trade-off between 
three desiderata: make plastic strains smooth, meet medical image observations,
and avoid using too much elastic energy (i.e., prefer to resolve shapes with plastic strains). 

\begin{figure}[!ht]
  \centering
  \includegraphics[width=1.0\hsize]{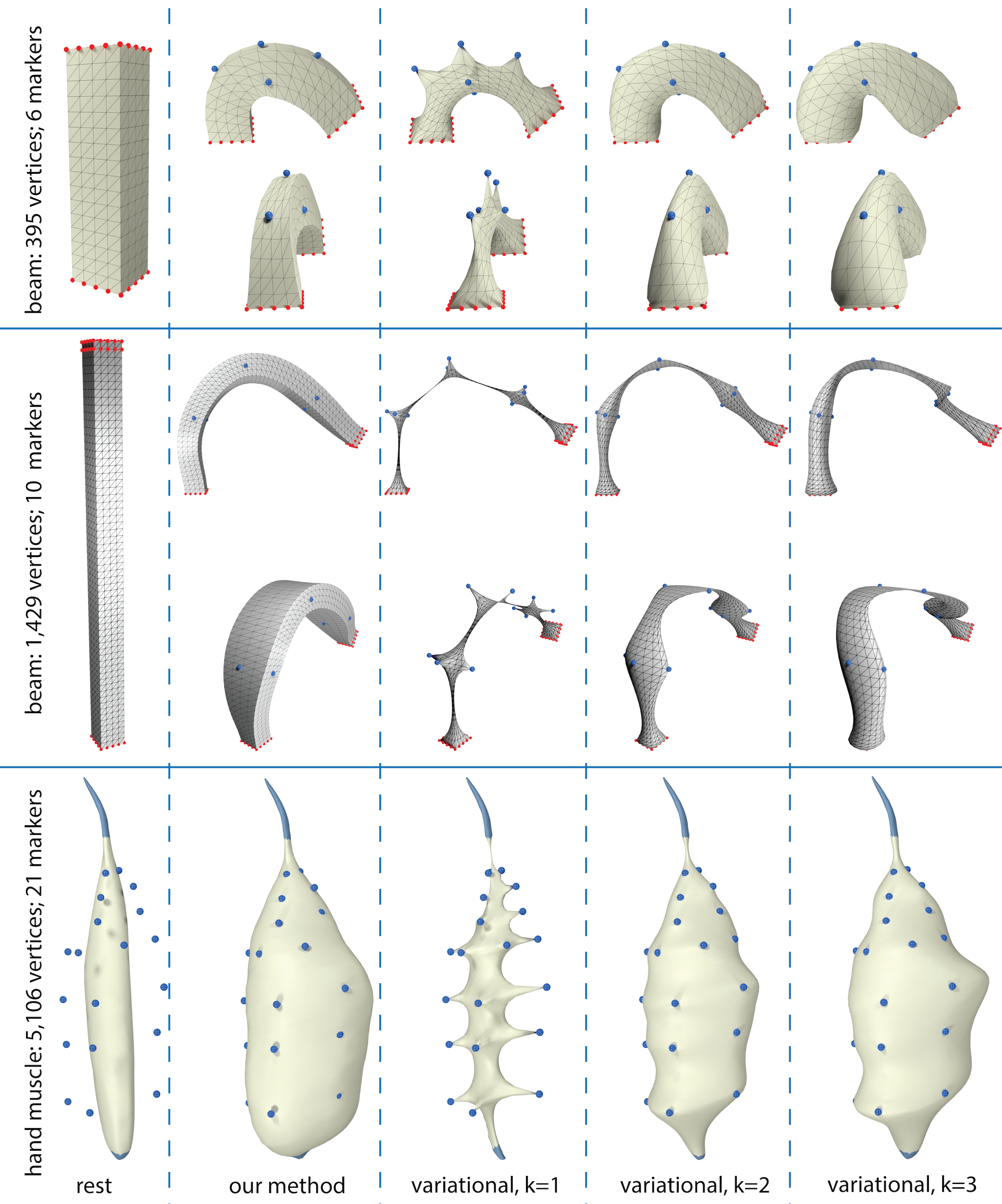}
  \figspace{}
  \vspace{-0.75cm}
  \caption{ {\bf Comparison to variational shape modeling.} 
  In a variational method~\cite{Botsch:2004:AIF}, 
  the wiggles increase if one imposes a stricter constraint satisfaction.
  First row: under a small number of landmarks, variational methods with $k=2,3$
  produce a smooth and reasonable result, albeit somewhat smoothening the rest shape.
  Middle row: under more landmarks, it becomes more difficult for variational methods
  to meet the landmarks while avoiding the wiggles.
  Bottom row: variational methods produce wavy results.
  Our method meets the landmarks and produces fewer wiggles.
  This is because the plastic deformation field can arbitrarily rotate and non-uniformly
  scale to adapt to the inputs; the elastic energy then finally irons out the kinks.
}
  \vspace{-0.25cm}
  \label{fig:kobbelt}
\end{figure}

Finally, we note that our formulation is different
to approaches that optimize the deformation gradient
$F$ directly (i.e., without an intermediary quantity such as the plastic deformation gradient). 
In Figures~\ref{fig:deformationTransfer} and~\ref{fig:kobbelt},
we compare to two such approaches: deformation transfer~\cite{Sumner:2004:DTF}
and variational shape modeling~\cite{Botsch:2004:AIF}.
We demonstrate that our method better captures
shapes defined using our inputs (landmarks, ICP markers, large spatially varying strains).
Among all compared approaches, the variational method in Figure~\ref{fig:kobbelt} came closest
to meeting our constraints, but there is still a 
visual difference to our method.
We provide a further comparison to variational methods in Section~\ref{sec:variational}.

\subsection{Solving the optimization problem for attached objects}
\label{sec:solveShapeOptimization}

\begin{figure*}[!t]
\centering
\includegraphics[width=1.0\hsize]{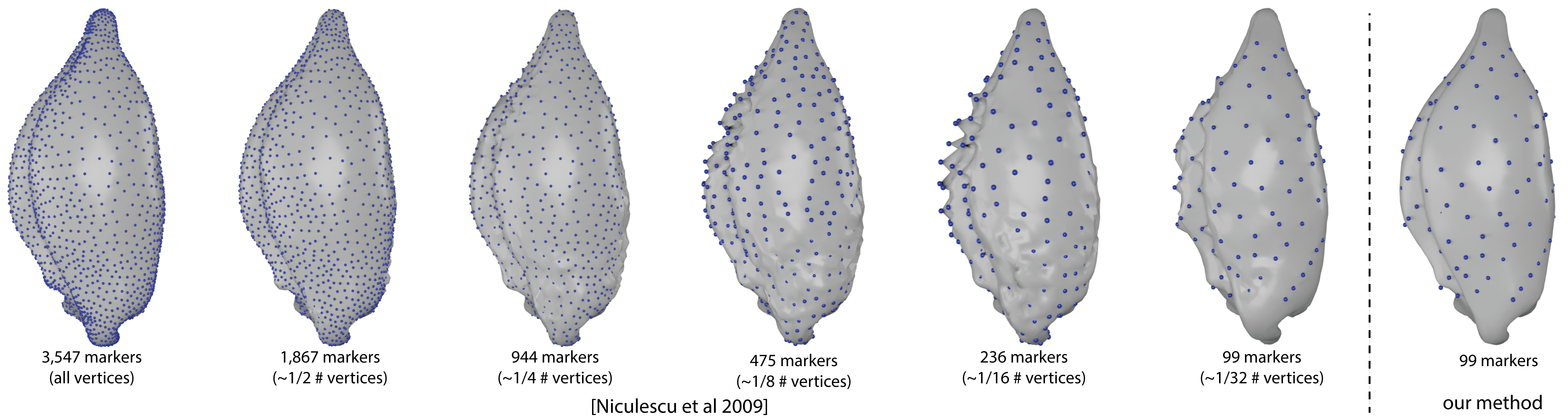}
\figspace{}
\vspace{-0.6cm}
\caption{ {\bf Elastic energy methods only work with dense markers.}
In this figure, we compare to a state-of-the art medical imaging technique~\cite{Niculescu:2009:ADM},
whereby the output shape is calculated by minimizing an elastic energy 
of a template shape, subject to dense medical image markers.
With dense markers, elastic energy methods work well (left). As the constraints
sparsify, elastic energy produces artifacts (middle).
Our plasticity method (right) produces a good shape even with sparse markers.
}
\vspace{-0.25cm}
\label{fig:medical}
\end{figure*}

We adapt the Gauss-Newton method~\cite{Sifakis:2005:ADF} to efficiently solve
the optimization problem of Equations~\ref{eq:objectiveFunction} 
and~\ref{eq:constraints} (example output shown in Figure~\ref{fig:musclesInMRI}). 
Doing so is not straightforward because a direct application of the Gauss-Newton method results in large dense matrices that are costly to compute and store, causing the method to fail on complex examples. Below, we demonstrate how to avoid these issues, producing a robust
method capable of handling complex spatially varying plastic strains. 
Before settling on our specific Gauss-Newton approach, we also attempted to use 
the interior-point optimizer available in the 
state-of-the-art Knitro optimization library~\cite{Knitro}. 
This did not work well because our problem is highly nonlinear.
The interior-point method (IPM) worked well on simple examples, but
was slow and not convergent on complex examples.
IPM fails because it requires the constraint Hessian, which is not available. When we approximated it, IPM generated intermediate states too far from the constraints, and failed. The strength of our Gauss-Newton approach is that we only need constraint gradients.
Our method inherits the convergence properties of the Gauss-Newton method. While not guaranteed to be locally convergent, Gauss-Newton is widely used because its convergence can approach quadratic when close to the solution. 

We note that our method is designed for sparse medical image landmarks and ICP markers.
In Figure~\ref{fig:medical}, we give a comparison to a related method from
medical imaging which used an elastic energy, but with dense correspondences.
Our method can produce a quality shape even under sparse inputs, and can consequently
work even with coarser MRI scans (such as our hip bone example; Figure~\ref{fig:hip-muscle}).
The ability to work with sparse markers also translates to
lower manual processing time to select the markers in the medical image.

Because the object is attached, Equation~\ref{eq:constraints}
implicitly defines $\BF{x}$ as a function of $\BF{s}.$
The Gauss-Newton method uses the
Jacobian $\BF{J}=d\BF{x} / d\BF{s},$ 
which models the change in the static equilibrium $\BF{x}$
as one changes the plastic deformation gradient. It eventually
relies on the derivative of elastic forces with respect
to the plastic deformation gradient, which we give 
in Appendix~\ref{app:appendixDerivatives}.
Because $\mathcal{E}_\textrm{MI}(\BF{x})$
and $\mathcal{E}_\textrm{a}(\BF{x})$ are quadratic functions 
of $\BF{x},$ we rewrite Equations~\ref{eq:objectiveFunction}
and~\ref{eq:constraints} as 
\begin{gather}
\label{eq:energy-full}
\argmin_{\BF{x},\BF{s}}
\frac{1}{2}||\BF{L} \BF{s}||^2  + 
\sum_{k=1}^{q+r+t} \frac{\BF{c_k}}{2} ||\BF{A_k}\BF{x}+\BF{b_k}||^2,\\
\label{eq:constraints0}
\text{s.t.}\quad \BF{f}_{\textrm{net}}(\BF{s},\BF{x})=\BF{0},
\end{gather}
where $\BF{f}_{\textrm{net}}(\BF{s},\BF{x})= 
\BF{f_\textrm{e}}\bigl(\BF{F_p}(\BF{s}), \BF{x}\bigr) + 
\BF{f_\textrm{a}}(\BF{x})$
is the net force on the mesh, and constant matrices, vectors and scalars $\BF{A_k}, \BF{b_k}, \BF{c_k}$ are independent 
of $\BF{s}$ and $\BF{x}$ (we give them in Appendix~\ref{app:Akbkck}).
The integer $t$ denotes the number of attachments.
We now re-write Equations~\ref{eq:energy-full} and~\ref{eq:constraints0} 
so that the plastic strains are expressed as $\BF{s} + \Delta \BF{s},$
and the equilibrium $\BF{x}$ as $\BF{x} + \Delta \BF{x},$ where
$\Delta \BF{x} = \BF{J} \Delta \BF{s}.$ 
At iteration $i$ of our Gauss-Newton method, given the previous iterates
$\BF{x}^i$ and $\BF{s}^i,$ we 
minimize a nonlinearly constrained  problem,
\begin{gather}
\label{eq:gn-iteration}
\argmin_{\BF{x}^{i+1},\Delta\BF{s}^i} \frac{1}{2}||\BF{L} (\BF{s}^i+\Delta\BF{s}^i)||^2+
\sum_{k=1}^{q+r+t} \frac{\BF{c_k}}{2} 
||\BF{A_k}\bigl(\BF{x}^i+\BF{J}\Delta\BF{s}^i\bigr)+\BF{b_k}||^2,\\
\label{eq:gn-constraint}
\text{s.t.}\quad
\BF{\BF{f}_{\textrm{net}}}(\BF{s}^{i}+\Delta\BF{s}^i, \BF{x}^{i+1})=\BF{0}.
\end{gather}
After each iteration, we update 
$\BF{s}^{i+1}=\BF{s}^{i}+\Delta\BF{s}^i.$
Observe that Equation~\ref{eq:gn-iteration} 
does not depend on $\BF{x}^{i+1},$
and that the constraint 
of Equation~\ref{eq:gn-constraint} is already differentially 
``baked'' into Equation~\ref{eq:gn-iteration} via $\Delta \BF{x} = \BF{J} \Delta \BF{s}.$
We therefore first minimize Equation~\ref{eq:gn-iteration} 
for $\Delta\BF{s}^i,$
using unconstrained minimization; call the solution 
$\overline{\Delta\BF{s}^i}.$
A naive minimization requires solving a 
large dense linear system of equations, 
which we avoid using the technique presented at the end of this section.
We regularize $\overline{\Delta\BF{s}^i}$ so that the 
corresponding $F_p$ is always positive-definite for each tet;
we do this by performing eigen-decomposition of the symmetric
matrix $F_p$ at each tet, and clamping any negative eigenvalues
to a small positive value (we use 0.01). Our method typically did not need
to perform clamping in practice, and in fact such clamping is usually
a sign that the method is numerically diverging, and should be restarted
with better parameter values.

We then minimize the optimization problem 
of Equations~\ref{eq:gn-iteration} and~\ref{eq:gn-constraint}
using a 1D line search, using the search direction 
$\overline{\Delta\BF{s}^i}.$
Specifically, for $\eta\geq 0,$
we first solve Equation~\ref{eq:constraints0} 
with $\BF{s}(\eta) := \BF{s}^i + \eta \overline{\Delta \BF{s}^i}$ 
for $\BF{x} = \BF{x}(\eta)$ 
using the Knitro library~\cite{Knitro}. Direct solutions
using a Newton-Raphson solver also worked, but we found Knitro
to be faster.
We then evaluate the objective
of Equation~\ref{eq:energy-full} at $\BF{x}=\BF{x}(\eta)$ 
and $\BF{s} = \BF{s}(\eta).$
We perform the 1D line search for the optimal $\eta$
using the Brent's method~\cite{Numerical:2007:NRT} because it does not require a gradient.

\paragraph{Initial guess:} We solve our optimization problem by first 
assuming a constant $s$ at each tet, starting from the template
mesh as the initial guess. This roughly positions, rotates and globally 
scales the template mesh to match the medical image. We use the output 
as the initial guess for our full optimization as described above. 

\begin{figure}[!ht]
  \centering
  \includegraphics[width=1.0\hsize]{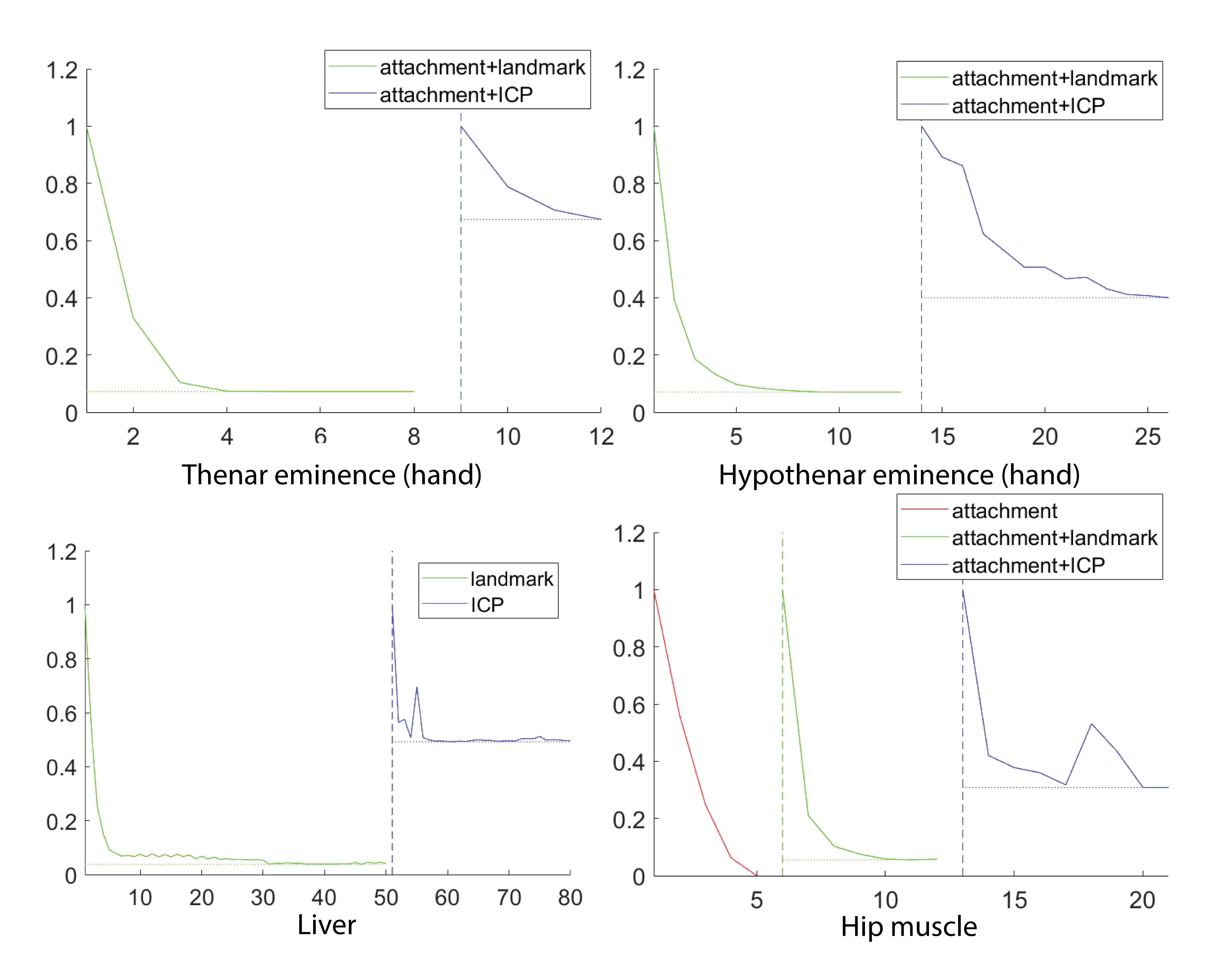}
  \figspace{}
  \vspace{-0.75cm}
  \caption{ {\bf Convergence plots.} 
X-axis are iterations, and Y-axis is the optimization energy.
The initial optimization energies are normalized to 1.0.
}
  \vspace{-0.25cm}
  \label{fig:convergencePlots}
\end{figure}

\paragraph{Optimization stages and stopping criteria:} 
We first do the optimization with attachments only. Upon convergence,
we add the landmarks, ignoring any ICP markers. This is because initially, 
the mesh is far away from the target and the ICP closest locations are unreliable.
After convergence, we disable the landmarks and enable the ICP markers
and continue optimizing. After this optimization meets a stopping criterium, we are done.
Our output is therefore computed with ICP markers only; landmarks only serve to guide the optimizer. 
This is because landmarks require a correct correspondence, 
and it is harder to mark this correspondence reliably in the scan than to simply 
select an ICP marker on the boundary of an organ. 
We recompute the closest locations to ICP markers after each Gauss-Newton iteration.
We stop the optimization if either of the following three criteria is satisfied:
(i) reached the user-specified maximal number of iterations (typically 20; but was
as high as 80 in the liver example), 
(ii) maximum error at ICP markers is less than a user-specified value 
(1mm for hand muscles),
(iii) the progress in each iteration is too small, determined by checking
if $\eta$ is under a user-defined threshold (we use 0.01).
Figure~\ref{fig:convergencePlots} shows the convergence of our optimization.

\paragraph{Avoiding dense large linear systems}
Because the object is attached, $\PP{\BF{f}_{\textrm{net}}}{\BF{x}}$ is 
square and invertible. Therefore, one can obtain a formula for $\BF{J}$ 
by differentiating 
Equation~\ref{eq:constraints0} with respect to $\BF{s},$
\begin{gather}
\label{eq:dxds-solve}
\BF{J}=-\Bigl(\PP{\BF{f}_{\textrm{net}}}{\BF{x}}\Bigr)^{-1}
\PP{\BF{f}_{\textrm{net}}}{\BF{s}}.
\end{gather}
The matrix $\BF{J}$ is dense (dimensions $3n\ \times\ 6m$).  
Observe that because Equation~\ref{eq:gn-iteration} is quadratic in 
$\Delta \BF{s}^i,$ minimizing it as done
above to determine the search direction $\overline{\Delta \BF{s}^i}$
is equivalent to solving a linear system with the system matrix $\BF{H},$
where $\BF{H}$ is the second derivative (Hessian matrix; dimension $6m\ \times\ 6m$) of 
Equation~\ref{eq:gn-iteration} with respect to $\Delta \BF{s}^i.$ 
Because $\BF{J}$ is dense, $\BF{H}$ is likewise a dense matrix,
\begin{gather}
\BF{H}=\BF{L}^2 + 
\BF{J}^T \bigl(\sum_{k} \BF{c_k}  \BF{A_k}^T \BF{A_k}\bigr) \BF{J}=\BF{L}^2+
\BF{Z}^T \BF{Z},\\
\label{eq:Z-equation}
\textrm{where}\quad \BF{Z}=\begin{bmatrix}
\BF{c_1} \BF{A_1} \\
\BF{c_2} \BF{A_2} \\
\vdots\\
\BF{c_r} \BF{A_r} 
\end{bmatrix}\BF{J}=-
\begin{bmatrix}
\BF{c_1} \BF{A_1} \\
\BF{c_2} \BF{A_2} \\
\vdots\\
\BF{c_r} \BF{A_{q+r+t}} 
\end{bmatrix} \Bigl(\PP{\BF{f}_{\textrm{net}}}{\BF{x}}\Bigr)^{-1}
\PP{\BF{f}_{\textrm{net}}}{\BF{s}}.
\end{gather}
Therefore, when the number of tet mesh elements $m$ is large,
it is not practically possible to compute $\BF{H},$ store it 
explicitly in memory or solve linear systems with it. 
To avoid this problem,
we first tried solving the system of equations using the 
Conjugate Gradient (CG) method.
This worked, but was very slow (Table~\ref{tab:H}).
The matrix $\BF{Z}\in\RR^{3(q+r+t)\, \times\, 6m}$ is\ dense.
In our complex examples, the number of medical image
constraints $q+r+t$ is small (typically 10\,-\,800) compared to the dimension 
of $\BF{s}$ ($6m$; typically \textasciitilde
200,000). Our idea is  to  efficiently compute 
the solution to a system $\BF{H} y = h$
for any right-hand side $h$ using the Woodbury matrix 
identity~\cite{Woodbury:1950:IMM}, where we view 
$\BF{L}^2$ as a ``base'' matrix and $\BF{Z}^T \BF{Z}$ 
a low-rank perturbation.
Before we can apply Woodbury's identity, we need to ensure that
the base matrix is invertible. 
As we prove in Appendix~\ref{app:laplacian},
the plastic strain Laplacian $\BF{L}$ is singular with  
six orthonormal vectors $\BF{\psi_i}$ in its 
nullspace (assuming that $\mathcal{M}$ is connected).
Each $\BF{\psi_i}$ is a vector of all ones in component $i$
of $s,$ $i=1,\ldots,6$ and all zeros elsewhere, 
divided by $\sqrt{m}$ for normalization.
It follows from the Singular Lemma (i) (Section~\ref{sec:singularLemma})
that $\BF{L}^{2}$ is also singular with the same  nullspace vectors. 
Therefore, we decompose 
\begin{equation}
\BF{H}=\Bigl(\BF{L}^2 - \sum_{i=1}^6\BF{\psi_i} \BF{\psi_i}^T\Bigr) + 
\Bigl(\BF{Z}^T \BF{Z} + \sum_{i=1}^6\BF{\psi_i} \BF{\psi_i}^T\Bigr) =
\BF{B} + \BF{\hat{Z}}^T \BF{\hat{Z}},
\end{equation}
where $\BF{B}=\BF{L}^2 - \sum_{i=1}^6\BF{\psi_i} \BF{\psi_i}^T$ and
$\BF{\hat{Z}}$ is matrix $\BF{Z}$ with an additional added 6 added rows 
$\BF{\psi_i}^T.$ By the Singular Lemma (iii) 
(Section~\ref{sec:singularLemma}),
$\BF{B}$ is now invertible, and we can use
Woodbury's identity to solve
\begin{gather}
y=\BF{H}^{-1}h=\Bigl(\BF{B}^{-1} - \BF{B}^{-1} \BF{\hat{Z}}^T 
\left(I+ \BF{\hat{Z}} \BF{B}^{-1} \BF{\hat{Z}}^T\right)^{-1} 
\BF{\hat{Z}} \BF{B}^{-1}\Bigr)h.
\end{gather}
We rapidly compute $\BF{Z},$ 
without ever computing or forming $\BF{J},$
 by solving sparse systems 
$\PP{\BF{f}_{\textrm{net}}}{\BF{x}}z = 
\BF{c_k} \BF{A_k}^T,$ for $k=1,\ldots,q+r+t.$ Observe that this sparse system matrix
is symmetric and the same for all $k.$ We factor it once using the Pardiso solver
and then solve the multiple right-hand sides in parallel.
The matrix $\BF{B}$ is constant,
and we only need to factor it once for the entire optimization. 
Finally, the matrix
$I+ \BF{\hat{Z}} \BF{B}^{-1} \BF{\hat{Z}}^T\in\RR^{3(q+r+t)\, \times\, 3(q+r+t)}$ 
is small, and so inverting it is fast. We analyze the 
performance of our algorithm in Table~\ref{tab:H}.

\begin{table}
\caption {\textbf{Solving a single linear system of equations with H,}
using the conjugate gradients and our  method. The naive direct solver failed in all cases. Note that $\BF{H}$ is a dense $6m\times 6m$ matrix. 
The column $t_{prep}$ gives a common pre-processing time for
both CG and our method.}
\label{tab:H}
\vspace{-0.3cm}
\begin{center}
\begin{tabular}{ l | c c | c | c c }
Example       & $6m$ & $3(q+r+t)$ & $t_{prep}$ & CG & Ours\\
\hline 
Hand muscle   & 237,954 & 1,143 & 17.5s & 897.5s & 9.5s \\
Hip bone      & 172,440 & 1,497 & 11.4s & 408.9s & 7.2s \\
Liver         & 259,326 & 1,272 & 24.4s & 1486.7s & 10.0s
\end{tabular}
\end{center}
\label{tab:fitting}
\vspace{-0.3cm}
\end{table}

\subsection{Singular lemma}
\label{sec:singularLemma}

\setlength{\columnsep}{9pt}
\begin{wrapfigure}[10]{r}{0.18\textwidth}
{\vspace{-0.4cm}} 
\includegraphics[width=0.18\textwidth]{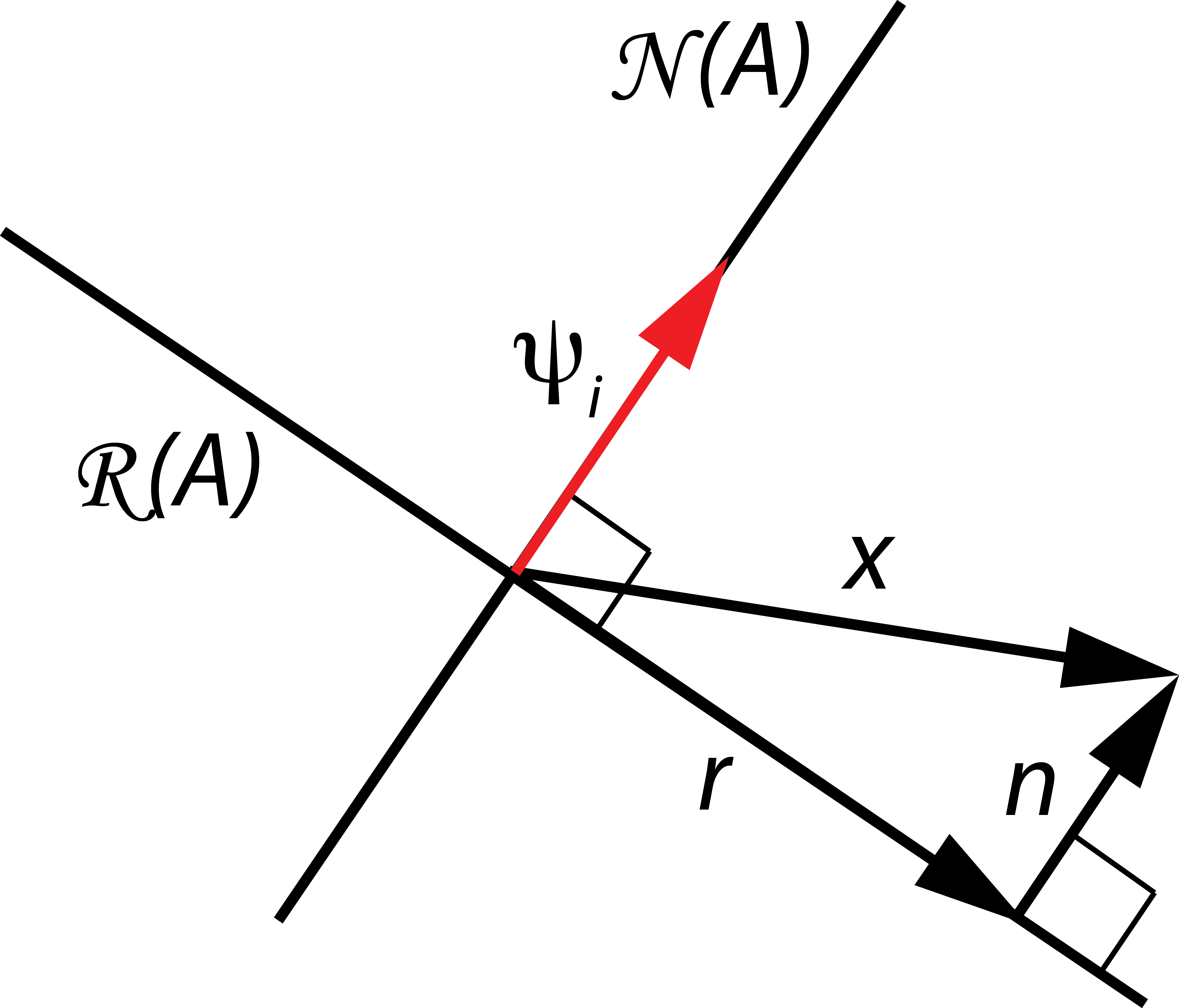}
\figspace{}
{\vspace{-0.6cm}}
\caption{
\textbf{Illustration of the Singular Lemma.}
}
\label{fig:singularLemma}
\end{wrapfigure}
In this paper, there are two occasions where we have 
to solve a singular sparse linear system with known nullspace
vectors. Such systems occur often in modeling of un-attached
objects, e.g., finding static equilibria, 
solving Laplace equations on the object's mesh,
animating with rotation-strain coordinates~\cite{Huang:2011:ISI}, 
or computing modal derivatives~\cite{Barbic:2005:RSI}. 
Previous work solved such systems ad-hoc, and the underlying theory has not been
stated or developed in any great detail. We hereby state and prove a lemma that comprehensively
surveys the common situations arising with singular systems in computer animation and simulation, 
and back the lemma with a mathematical proof (Appendix~\ref{app:singularLemma}).
Recall that the \emph{nullspace} of a matrix $A\in\RR^{p\times p}$ 
is $\mathcal{N}(A)=\{x\in\RR^p\,\, ;\,Ax=0\}$, 
and the \emph{range} of $A$ is $\mathcal{R}(A)=\{Ax\,;\,x\in\RR^p\}.$
Both are linear vector subspaces of $\RR^p.$\\

\noindent \textbf{Singular Lemma:}
Let the square symmetric matrix $A\in\RR^{p\times p}$ be singular with
a known nullspace spanned by $k$ linearly independent 
vectors $\psi_1,\ldots,\psi_k.$ Then the following statements hold:\\
(i) $\mathcal{N}(A)$ and $\mathcal{R}(A)$ are orthogonal.
Every vector $x\in\RR^p$ can be uniquely expressed as 
$x=n+r,$ where $n\in\mathcal{N}(A)$ 
and $r\in\mathcal{R}(A).$ Vector $r$ is orthogonal to $n$
and to $\psi_i$ for all $i=1,\ldots,k$ (Figure~\ref{fig:singularLemma}).
\\
(ii) Let $b\in\mathcal{R}(A).$ Then, the singular system $Ax=b$ has
a unique solution $x$
with the property that $x$ is orthogonal 
to $\psi_i$ for all $i=1,\ldots,k.$
This solution can be found by solving the \emph{non-singular} linear system
\begin{equation}
\begin{bmatrix}
A & \psi_1 & \ldots & \psi_k \\
\psi_1^T & 0 &\ldots & 0\\
\vdots   & \vdots  & & \vdots \\
\psi_k^T & 0 &\ldots & 0
\end{bmatrix}
\begin{bmatrix}
x \\
\lambda_1 \\
\vdots   \\
\lambda_k 
\end{bmatrix}
=
\begin{bmatrix}
b \\
0 \\
\vdots   \\
0 
\end{bmatrix}.
\label{eq:singularSolve}
\end{equation}
All other solutions equal  $x+\sum_{i=1}^k \mu_i \psi_i$
for some scalars $\mu_i\in\RR.$\\
(iii) For any scalars $\alpha_i\ne 0,$ the matrix
$B=A+\sum_{i=1}^k \alpha_i \psi_i \psi_i^T$ is invertible. If 
$\psi_i$ are orthonormal vectors, then
the solution to 
$By=h$ equals  $y=x+\sum_{i=1}^k \frac{\lambda_i}{\alpha_i} \psi_i,$
where $x$ and $\lambda_i$ are solutions to Equation~\ref{eq:singularSolve}
with $b=\textrm{proj}_{\mathcal{R}(A)}h=h-
\sum_{i=1}^k (\psi_i^T h)\psi_i.$
We give the proof of the singular lemma in Appendix~\ref{app:singularLemma}.

\subsection{Un-attached objects}
\label{sec:unattached}

\begin{figure*}[!ht]
\centering
\includegraphics[width=0.9\hsize]{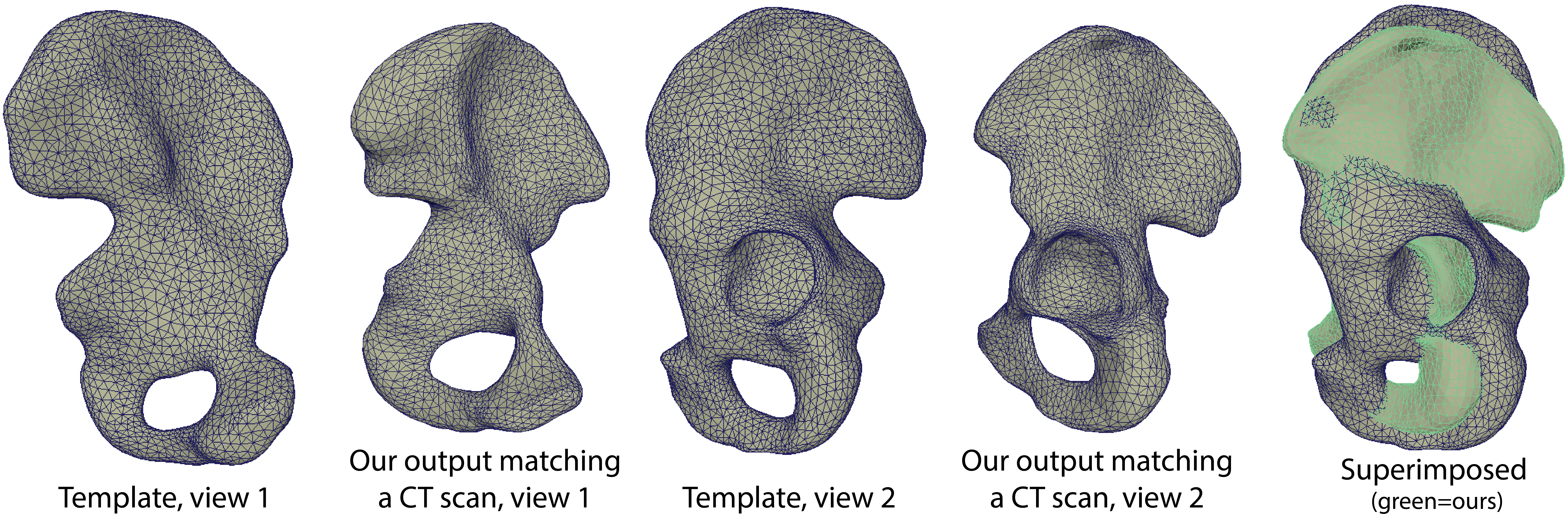}
\figspace{}
\vspace{-0.3cm}
\caption{ {\bf Un-attached optimization of a hip bone shape to a CT scan.} 
The scanned bone is smaller and has a substantially different shape to the template.  
}
\vspace{-0.25cm}
\label{fig:hipBone}
\end{figure*}

\begin{figure*}[!ht]
\centering
\includegraphics[width=0.9\hsize]{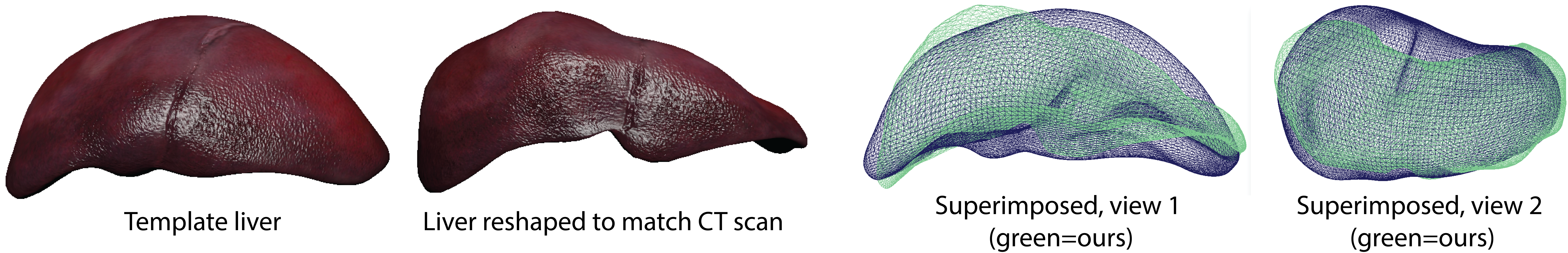}
\figspace{}
\vspace{-0.5cm}
\caption{ {\bf Un-attached optimization of a liver shape to a CT scan.} 
Our method successfully captures the large shape variation
between the template and the scan. 
This figure also demonstrates that our method makes it possible to transfer 
the rendering textures and uv coordinates from the template onto the output.}
\vspace{-0.25cm}
\label{fig:liver}
\end{figure*}

\begin{figure*}[!ht]
        \centering
        \includegraphics[width=0.9\hsize]{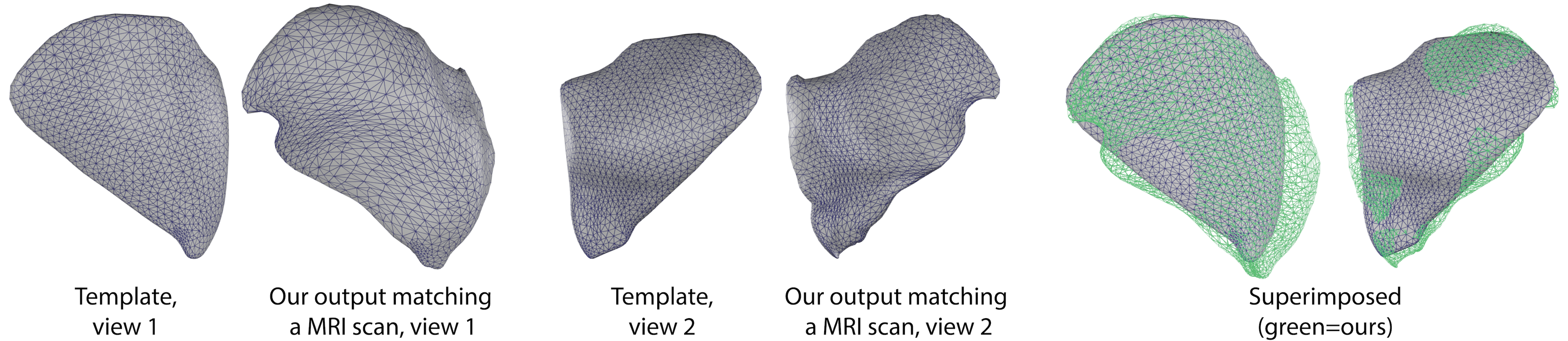}
        \figspace{}
        \vspace{-0.3cm}
        \caption{ {\bf Attached optimization of a hip muscle (gluteus medius) to a MRI scan.} 
                Our method successfully captures the large shape variation
                between the template and the scan.}
        \vspace{-0.25cm}
        \label{fig:hip-muscle}
\end{figure*}

The difficulty with un-attached objects is that 
we now have $\BF{f}_{\textrm{net}} =\BF{f_\textrm{e}},$
and the equation
$\BF{f_\textrm{e}}\bigl(\BF{F_p}(\BF{s}), 
\BF{x}\bigr)=0$
no longer has a unique solution $\BF{x}$ 
for a fixed plastic state $\BF{s}.$
This can be intuitively easily understood: one can arbitrarily
translate and rotate any elastic equilibrium
shape $\BF{x}$ under the given plastic state $\BF{s};$
doing so produces another elastic equilibrium shape. The
space of solutions $\BF{x}$ is 6-dimensional.
This means that we can no longer uniquely solve
Equation~\ref{eq:constraints0} for $\BF{x}$ during our line search
of Section~\ref{sec:solveShapeOptimization}.
Furthermore, the square tangent stiffness matrix
\begin{equation}
\BF{K}(\BF{F_p}(\BF{s}),\BF{x})=
\PP{\BF{f_\textrm{e}}
\bigl(\BF{F_p}(\BF{s}),\BF{x}\bigr)}{\BF{x}}
\end{equation}
is no longer full rank. 
In order to address this, we now state and prove the following Nullspace Lemma.\\\smallskip

\noindent \textbf{Nullspace Lemma:}
The nullspace of the tangent stiffness matrix of an elastoplastic deformable object
in \emph{static equilibrium} $\BF{x}$ under plasticity, is 6-dimensional.
The six nullspace vectors are
$\psi_i := [e_i,e_i,\ldots,e_i],$
where $e_i\in\RR^3$ is the $i$-th
standard basis vector, and
$\psi_{3 + i} := [e_i\times x_1, e_i\times x_2, 
\ldots, e_i\times x_n],$ for $i=1,2,3.$

 To the best of our knowledge, this fact of elasto-plasto-statics
has not been stated or proven in prior work. 
It is very useful when modeling large-deformation elastoplasticity, 
as real objects are often un-attached, or attachments cannot be 
easily modeled. We give a proof in Appendix~\ref{app:nullspaceLemma}.
To accommodate un-attached objects, it is therefore 
necessary to stabilize the translation and rotation. 
For translations, this could be achieved easily by 
fixing the position of any chosen vertex. 
Matters are not so easy for rotations, however. 
Our idea is to constrain the centroid 
of all tet mesh vertices to a specific given position $t,$
and to constrain the ``average rotation'' of the model
to a specific given rotation $R.$ We achieve this using 
the familiar ``shape-matching''~\cite{Mueller:2005:MDB},
by imposing that the rotation in  the polar decomposition
of the global covariance matrix must be $R.$
We therefore solve the following optimization problem,
\begin{gather}
\label{eq:energy-full1}
\argmin_{\BF{x},\BF{s},R,t}\ ||\BF{L}\, \BF{s}||^2 + 
\alpha \mathcal{E}_\textrm{MI}(\BF{x})\\
\label{eq:constraints1}
\textrm{s.t.}\quad 
\BF{f_\textrm{e}}\bigl(\BF{F_p}(\BF{s}),\BF{x}\bigr)=\BF{0},\\
\label{eq:constraints2} 
\left(\sum_{j \in D} w_j x_j\right) - \sum_{j \in D} w_j X_j = t,\\
\label{eq:constraints3} 
\mathrm{Polar}\left(\sum_{j \in D} 
w_j \Bigl(x_j -t\Bigr)
\Bigl(X_j - \bigl(\sum_{k \in D} w_k X_k\bigr)\Bigr)^T\right) = R,
\end{gather}
where $D$ is the set of 
points on the mesh surface where we have either 
a landmark or an ICP constraint, $w_j$ is the weight of a point,
$X_j$ is the position of vertex $j$ in $\mathcal{M}$ 
and $\mathrm{Polar}(F)$ is the polar decomposition 
function that extracts the rotational part of  a  matrix $F.$
We set all weights equal, i.e., $w_j=1/|D|.$
We choose the set $D$ as opposed to all mesh vertices
so that we can easily perform optimization with
respect to $R$ and $t$ (next paragraph).
We assume that our argument matrices $F$ to $\mathrm{Polar}$ are
not inversions, i.e., $\textrm{det}(F)>0,$
which establishes that $\mathrm{Polar}(F)$ is always a rotation
and not a mirror. 
This requirement was easily satisfied in our examples, and is
essentially determined by the medical imaging constraints;
the case $\textrm{det}(F)<0$ would correspond to an inverted
(or mirror) medical image, which we exclude.

We solve the optimization problem of  
Equations~\ref{eq:energy-full1},
~\ref{eq:constraints1},~\ref{eq:constraints2} 
and~\ref{eq:constraints3}
using a block-coordinate descent, by iteratively optimizing
$\BF{x},\BF{s}$ while keeping $R,t$ fixed and vice-versa
(Figure~\ref{fig:hipBone},~\ref{fig:liver}). 
Rigid transformations do not affect  smoothness of $\BF{s}$ 
so we do not need to consider  it when optimizing $R, t.$ 
We need to perform two modifications to our Gauss-Newton iteration 
of Section~\ref{sec:solveShapeOptimization}. 
The first modification is that we need to simultaneously
solve Equations~\ref{eq:constraints1},~\ref{eq:constraints2} 
and~\ref{eq:constraints3} when determining the static equilibrium
in the current plastic state $\BF{s}.$ As with attached objects,
we do this using the Knitro optimizer. In order to do this, 
we need to compute the first
and second derivatives of the stabilization constraints in 
Equations~\ref{eq:constraints2} and~\ref{eq:constraints3} 
(Section~\ref{sec:polarGradient}). The second modification is 
needed because the tangent stiffness matrix
$\BF{K}(\BF{F_p}(\BF{s}),\BF{x}),$ as explained above, is 
now singular with a known 6-dimensional nullspace. 
In order to compute the Jacobian matrix $\BF{J}$
using Equation~\ref{eq:dxds-solve}, we use our Singular Lemma (ii) 
(Section~\ref{sec:singularLemma}). Note that the right-hand side is 
automatically in the range of $\BF{K}$ because 
 Equation~\ref{eq:dxds-solve} was obtained by differentiating
a valid equation, hence Equation~\ref{eq:dxds-solve} must
also be consistent. 

\subsection{Gradient and Hessian of $\mathrm{Polar}(F)$}
\label{sec:polarGradient}

Previous work computed first and second-order \emph{time} derivatives of the rotation matrix
in polar decomposition~\cite{Barbic:2011:RLS}, or
first derivative with respect to each individual entry 
of $F$~\cite{Twigg:2010:PCG,Chao:2010:ASG}. In our work, we need  
the first \emph{and} second derivatives of $R$ with respect 
to each individual entry of $F.$ We found an elegant approach
to compute them using Sylvester's equation, as follows.
Observe that $\mathrm{Polar}(F) = F S^{-1},$
where $S$ is the symmetric matrix in the polar decomposition.  
Because $\textrm{det}(F) > 0,$ $S$ is positive-definite
and uniquely defined as $S=\sqrt{F^TF}.$ 
To compute the first-order derivatives, we start from $F=RS$,
and differentiate,\begin{gather}
\PP{F}{F_i}=\PP{R}{F_i}S+R\PP{S}{F_i},\quad \textrm{hence}\quad
\PP{R}{F_i}=\left(\PP{F}{F_i}-R\PP{S}{F_i}\right)S^{-1},
\end{gather}
Therefore, we need to compute $\PPI{S}{F_i}.$ We have
\begin{gather}
F^TF=S^2,\quad \textrm{and\ thus\ }\quad
\label{eq:S-1st-deriv}
\PP{F^TF}{F_i}=\PP{S}{F_i}S+S\PP{S}{F_i},
\end{gather}
i.e., this is the classic Sylvester equation 
for the unknown matrix $\PP{S}{F_i}$~\cite{Sylvester:1884:SLE}. 
The Sylvester equation $AX+XB=C$
can be solved as 
\begin{equation}
\left(B^T \oplus A\right)^{-1} \text{vec}(X)=\text{vec}(C),
\end{equation}
where $\oplus$ is the Kronecker sum of two matrices. In our case,
\begin{equation}
\text{vec}(\PP{S}{F_i})=\left(S \oplus S\right)^{-1}\text{vec}(\PP{F^TF}{F_i}).
\end{equation}
The computation of second-order derivatives follows the same
recipe: differentiate the polar decomposition and solve a Sylvester
equation. We give it in Appendix~\ref{app:secondPolar}.

We can now compute the gradient and Hessian of our stabilization
constraints. The translational constraint is linear in  $\BF{x}$ and 
can be expressed as $W_1 \BF{x} - d_1=0,$ 
where $W_1$ is a $3\times 3n$ sparse matrix. 
Although $\mathrm{Polar}$ is not linear, 
the argument of $\mathrm{Polar}$ is linear in $\BF{x}.$ 
The rotational constraint can be expressed as 
$\mathrm{Polar}\left(W_2 \BF{x} - d_2\right)-\bar{R}=0,$
where $W_2$ is a $9\times 3n$ sparse matrix.
The Jacobian of the translational constraint
is $W_1,$ and the Hessian is zero. 
For the rotational constraint, the Jacobian 
is $\PP{R}{F}
\colon W_2 \in \RR^{9\times 3n}$ 
and the Hessian is $(W_2^T \colon \PPP{R}{F} \colon
W_2)\in \RR^{9\times 3n \times 3n},$ where $:$
denotes tensor contraction.

\begin{figure}[!ht]
\centering
\includegraphics[width=1.0\hsize]{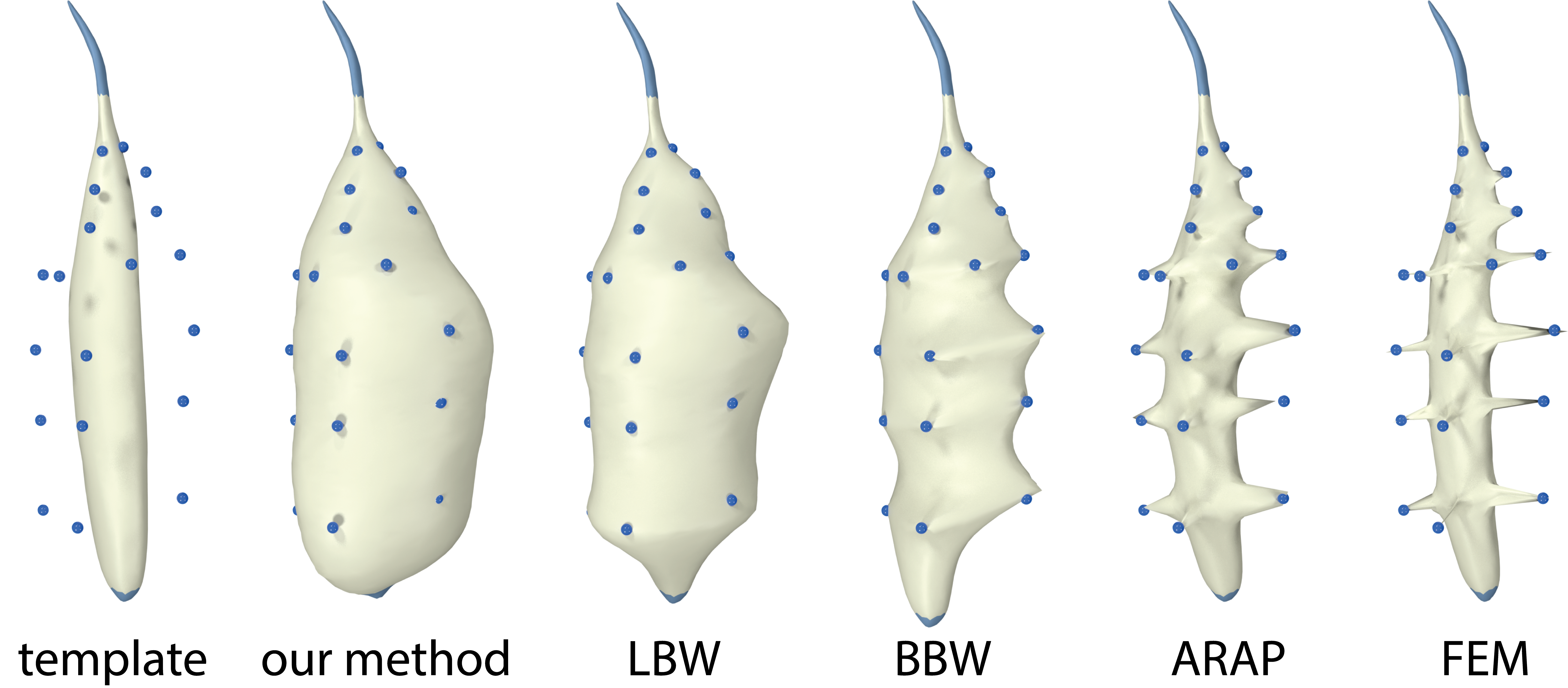}
\figspace{}
\vspace{-0.6cm}
\caption{ {\bf Standard volume-based shape deformation methods result in wiggly and spiky artefacts.}
The shown hand Palmar interossei muscle has a tendon on one side and no tendon 
on the other; both ends are attached to a bone (light blue).
Definitions of the acronyms are in Section~\ref{sec:standardShapeDeformation}.
The MRI landmark constraints are shown in dark blue.
The shape deformation between the template muscle and the shape in 
MRI consists of large and spatially varying stretches.
Our method successfully models this deformation.
We note that spikes cannot be avoided by using, say, a spherical region
for the constraints as opposed to a point;
the non-smoothness just moves to the spherical region boundary.
}
\vspace{-0.25cm}
\label{fig:ARAP}
\end{figure}

\begin{figure}[!ht]
\centering
\includegraphics[width=1.0\hsize]{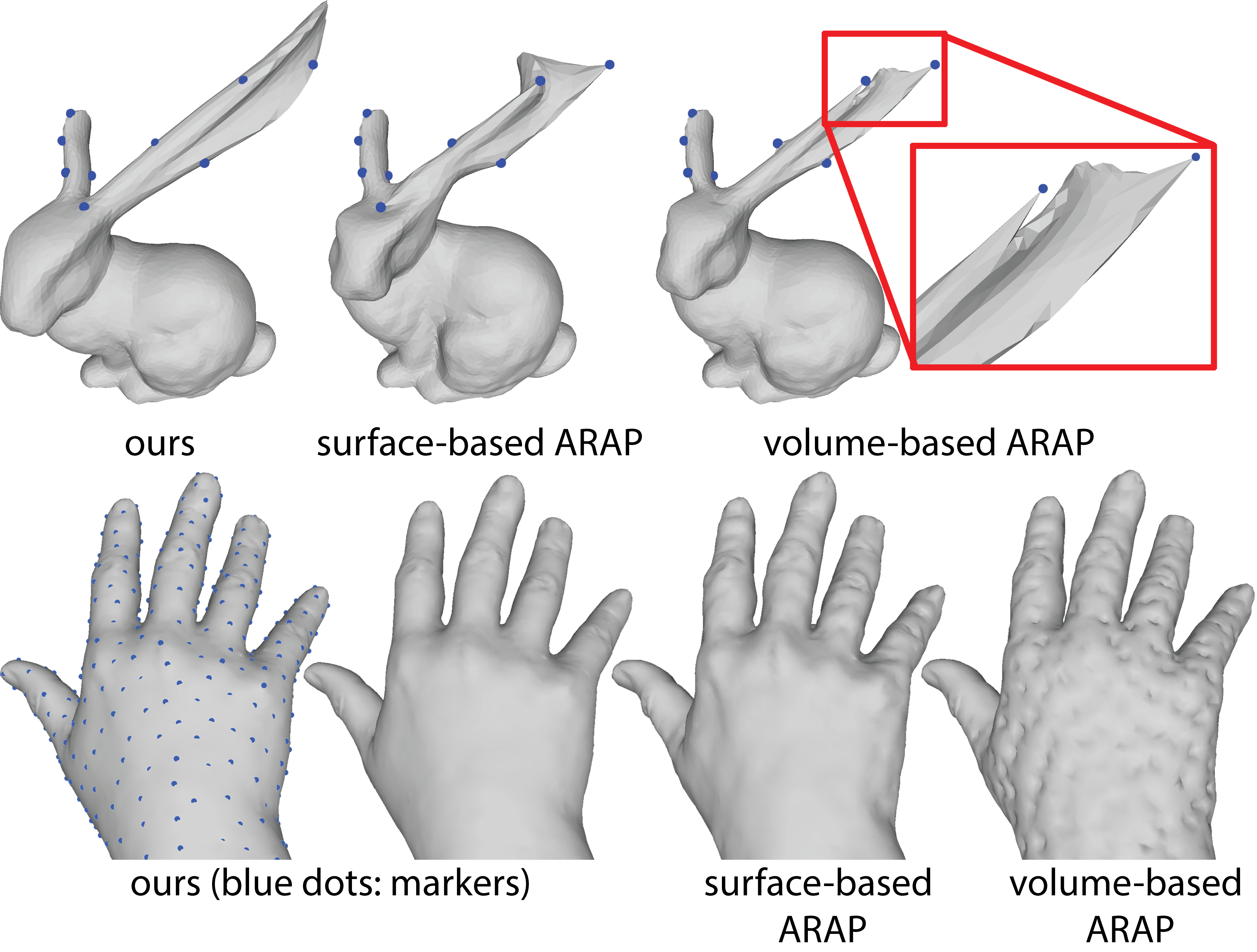}
\figspace{}
\vspace{-0.6cm}
\caption{ {\bf Comparison between a surface energy and volumetric energy.}
In both examples (bunny and hand), we performed non-rigid iterative closest point alignment between a template triangle mesh, and a collection of target ICP markers (13 for bunny and 456 for the hand). For the bunny, we manually placed the markers to greatly enlarge bunny's left ear. For the hand, we placed the markers on a pre-existing target hand surface mesh that has different geometric proportions and mesh topology as the template. Template is a man and target is a woman. We then solved the ICP problem using the surface-based ARAP\ energy, volume-based ARAP energy and our volumetric plastic strains. Our method produces smooth artifact-free outputs.}
\vspace{-0.25cm}
\label{fig:spikyArap}
\end{figure}

\subsection{Comparison to standard shape deformation methods}
\label{sec:standardShapeDeformation}

Our shape deformation setup is similar to 
standard shape deformation problems
in computer graphics. In fact, we first attempted to solve the 
shape deformation problem with 
as-rigid-as-possible energy (ARAP)~\cite{Sorkine:2007:ARA},
bounded biharmonic weights (BBW)~\cite{Jacobson:2011:BBW},
biharmonic weights with linear precision (LBW)~\cite{Wang:2015:LSD},
and a Finite Element Method static solver (FEM)~\cite{Barbic:2009:DOA}.
Unfortunately, none of the methods worked well. 
Figures~\ref{fig:ARAP} and~\ref{fig:spikyArap} demonstrate that these methods 
produce non-smooth shapes with spikes (ARAP, BBW, FEM), or wiggles (LBW). 

Mathematically, the reason for the spikes in ARAP, BBW and FEM is that
point constraints for second-order methods are inherently flawed.
As one refines the solution by adding more tetrahedra, the 
solution approaches a spiky point function at each point constraint,
which is obviously not desirable.
This mathematical issue is exposed in our work because
our shape deformation consists of large spatially varying stretches. 
Often, the template mesh needs to be stretched $\sim$2x or more along
some (or several) coordinate axes. The medical image constraints are 
distributed all around the muscle, pulling in different directions
and essentially requesting the object to undergo a spatially
non-uniform and anisotropic scale. This exacerbates the spikiness for second-order methods.
We note that these problems cannot be avoided simply by using an elastic
energy that permits volume growth.
Namely, Drucker's stability condition~\cite{Drucker:1957:DSI} 
requires a monotonic elastic energy increase with increase in strain. 
An elastic energy therefore must penalize strain increases
if it is to be stable; and this impedes large-strain modeling in methods
that rely purely on an elastic energy. Our plasticity method does not penalize large strains 
and thus avoids this problem. Spikes can be avoided by using a higher-order variational
method such as LBW. However, our experiments indicate that such methods suffer from wiggles
when applied to medical imaging problems (see also Figures~\ref{fig:kobbelt} and~\ref{fig:variational}).

\secspace
\section{Results}
\secspace
\label{sec:results}

\begin{figure}
        \centering
        \includegraphics[width=1.0\hsize]{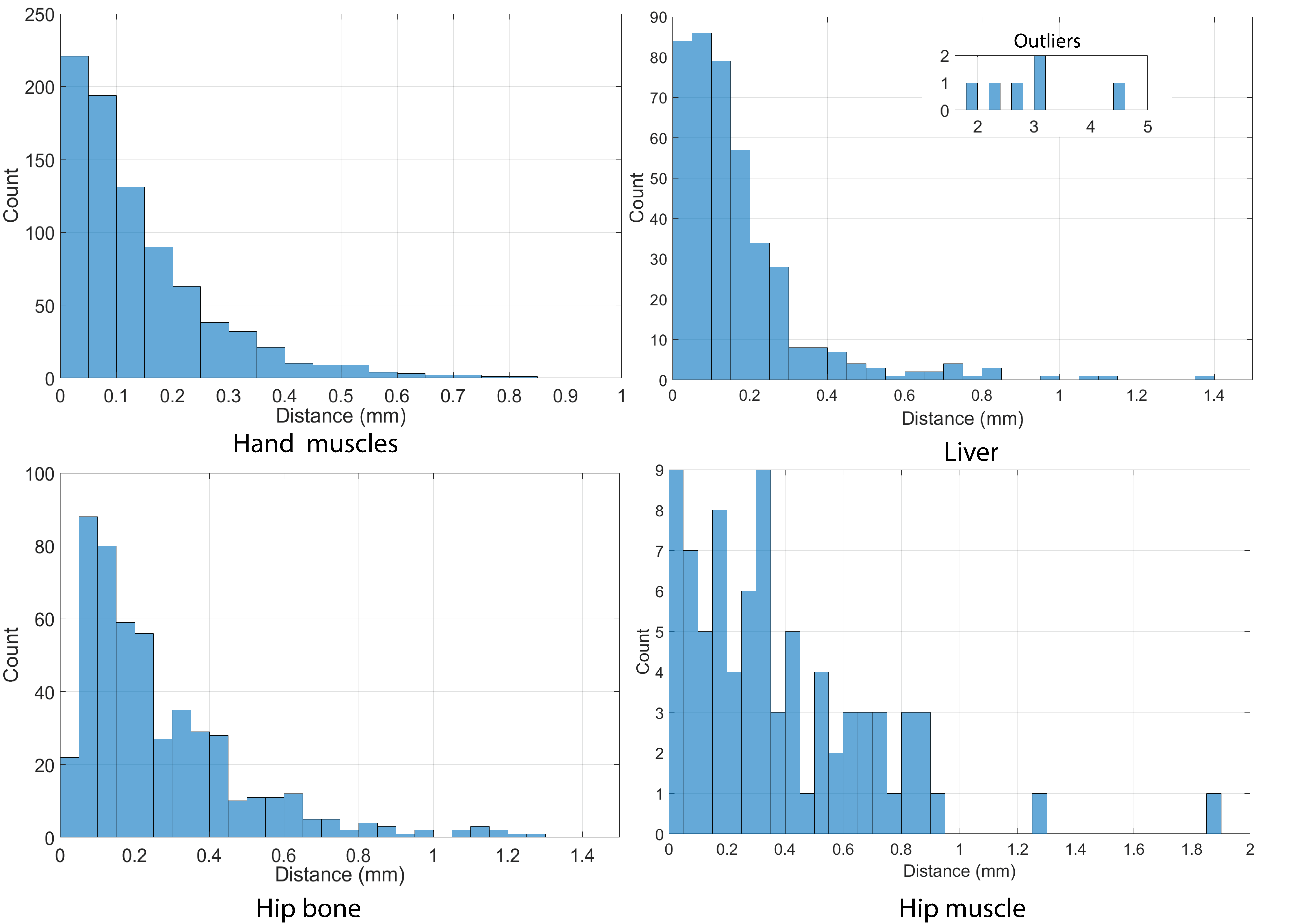}
        \figspace{}
        \vspace{-0.75cm}
        \caption{ {\bf Output ICP error histograms.} 
                Each medical image marker contributes 1 entry to the histogram. 
                The hand muscles histogram is a combined histogram for all the 17 hand muscles.}
        \vspace{-0.25cm}
        \label{fig:histograms}
\end{figure}

\begin{figure}
        \centering
        \includegraphics[width=1.0\hsize]{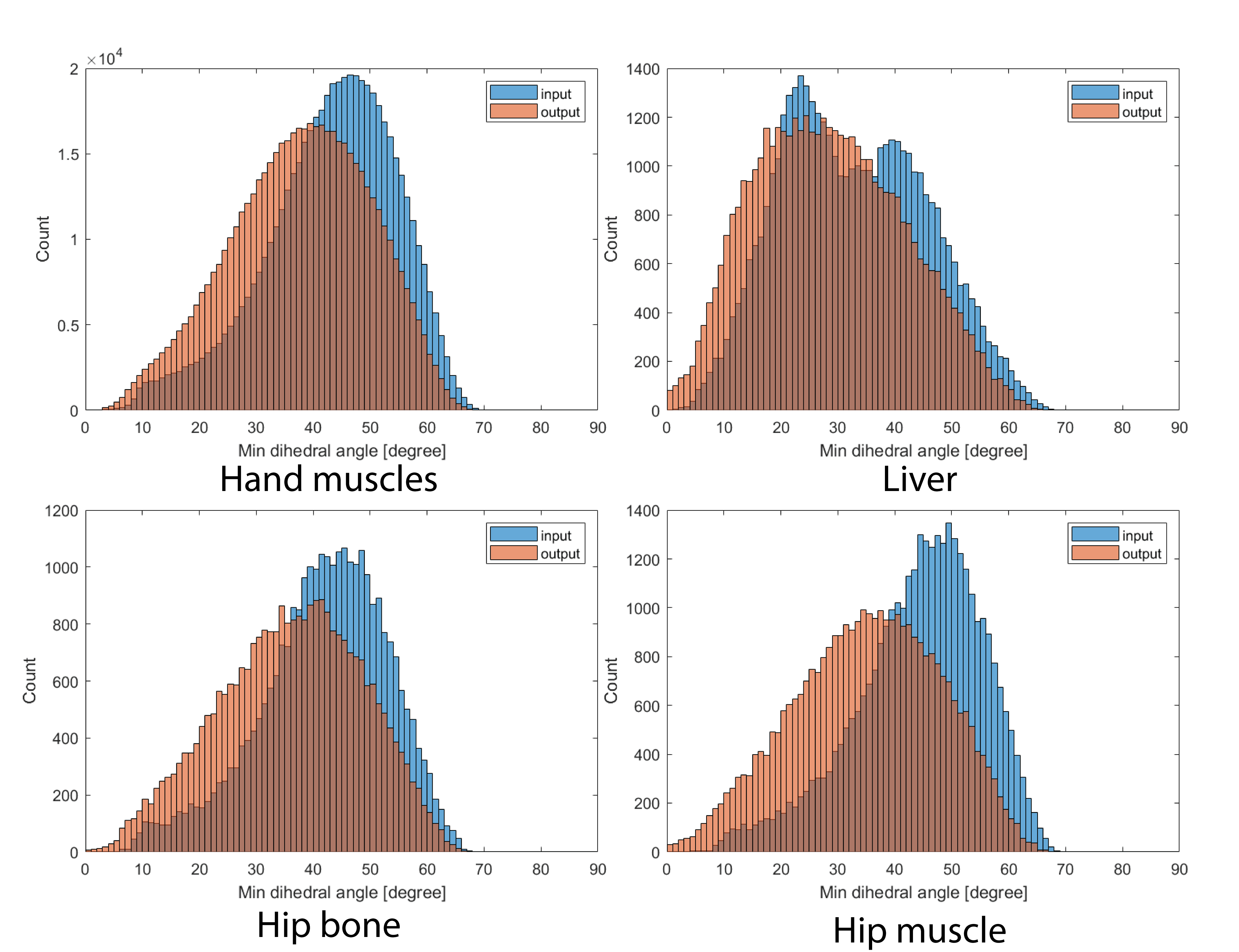}
        \figspace{}
        \vspace{-0.75cm}
        \caption{ {\bf Minimum tetrahedral dihedral angles before and after our optimization.} 
    It can be seen that the output angles are still relatively large.
    As expected, the output angles are somewhat worse than the initial ones, as the object
    has undergone a plastic deformation.
}
        \vspace{-0.25cm}
        \label{fig:minDihedralAngles}
\end{figure}

\begin{figure}
\centering
\includegraphics[width=0.9\hsize]{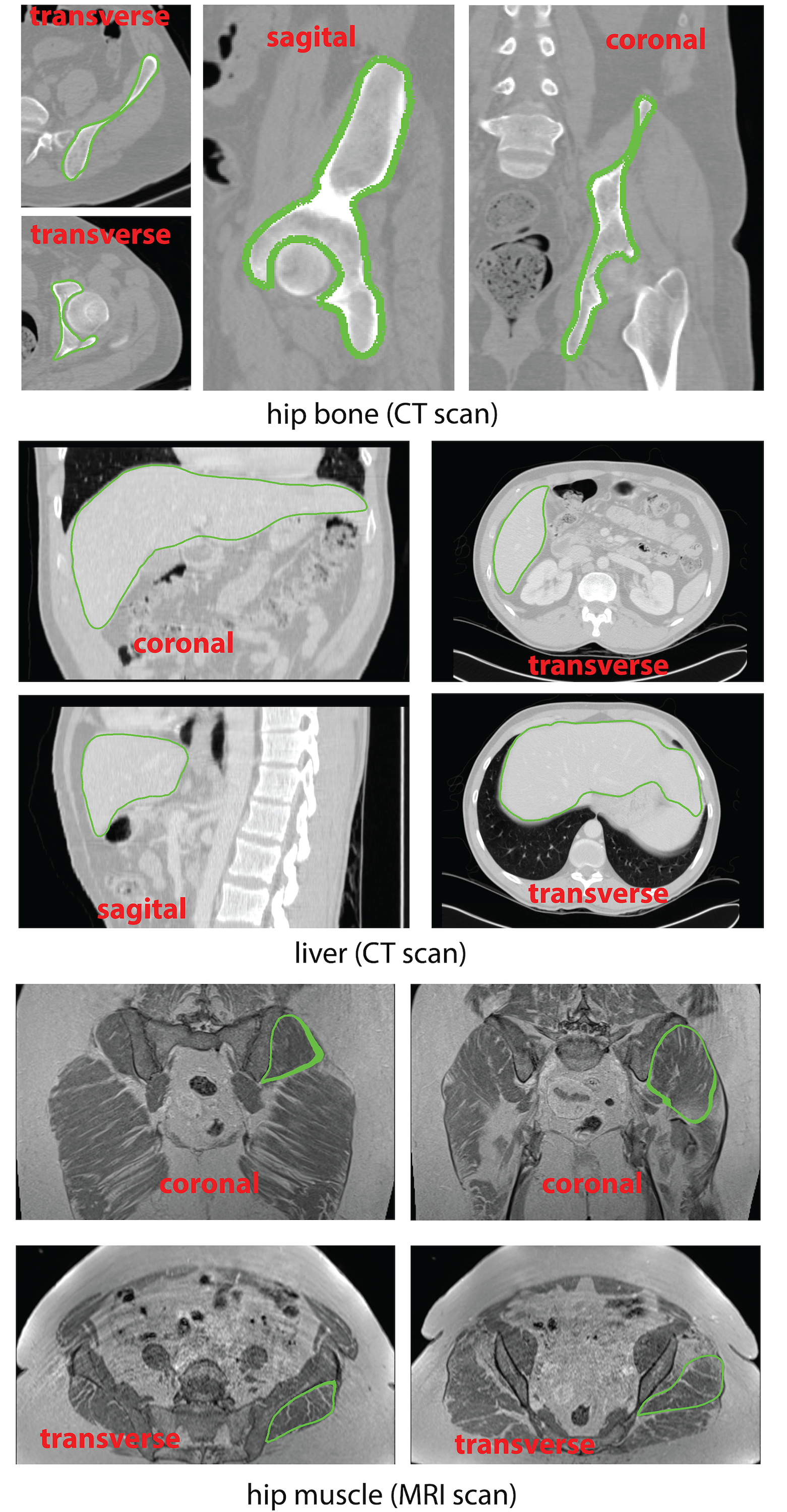}
\figspace{}
\vspace{-0.4cm}
\caption{ {\bf Our extracted organs match the medical image.} 
The intersection of the output mesh with the medical image slices is shown in green.
}
\vspace{-0.25cm}
\label{fig:slices}
\end{figure}

We extracted muscles of the human hand and the hip muscle from an MRI
scan, and a hip bone and a liver from a CT scan.
We analyze the performance of our method
in Table~\ref{tab:time}. 
In Figure~\ref{fig:histograms},
we give histograms of the
magnitude of the difference between the positions of the medical image
markers and their output positions.
It can be seen that our method  produces
shapes that generally match the medical image constraints to
0.5mm or better.
In Figure~\ref{fig:minDihedralAngles}, we demonstrate that the output quality
of our tetrahedra is still good; if needed, this could be further 
improved by re-meshing~\cite{Bargteil:2007:FEM}.
Figures~\ref{fig:slices} and~\ref{fig:removeInterpenetrations} superimpose our output meshes 
on the CT and MRI scans, respectively.
In Figure~\ref{fig:vipss}, we compare to a recent
implicit point set surface reconstruction method~\cite{Huang:2019:VIP}.
In Figure~\ref{fig:plasticGroundTruth}, we evaluate our method in the
presence of known ground truth plastic deformation gradients.
Figure~\ref{fig:surfaceMethods} provides a comparison to surface-based
methods. 

\begin{table*}[t]
\caption {\textbf{The statistics for our examples:} 
\#vtx=number of vertices; \#ele = number of tetrahedra in $\mathcal{M};$
\#iter=number of ICP iterations; ``time''=total computation time to compute the output shape; 
``attached'' means whether the object is attached or not; 
$e_\textrm{init}$ = error between our template mesh and the ICP constraints; 
$e_\textrm{final}$ = error between our result and the ICP markers.
The first and second reported error numbers are the average and maximum error, respectively.
In the hand example, there are 17 groups of muscles;
``min'', ``med'' and ``max'' refers to the smallest, representative median and largest muscle group;
``max-m'' is the example with the largest number of ICP constraints.}
\label{tab:time}
\vspace{-0.25cm}
\begin{center}
\begin{tabular}{ l | c c c | c c | c c c }
Example       & \# vtx & \# ele & \# markers & \# iter & time [min] & attached & $e_\textrm{init} [mm]$ & $e_\textrm{final} [mm]$ \\
\hline 
Hand muscle (min)   & 4,912  & 20,630  & 15  & 8  & 14.2  & yes & 0.53\,/\,2.22  & 0.06\,/\,0.14 \\
Hand muscle (med)   & 6,260  & 31,737  & 32  & 12 & 12.8  & yes & 0.62\,/\,1.56  & 0.11\,/\,0.55 \\
Hand muscle (max)   & 8,951  & 42,969  & 96  & 11 & 14.2  & yes & 3.35\,/\,12.82 & 0.11\,/\,0.34 \\
Hand muscle (max-m) & 7,552  & 34,966  & 151 & 18 & 20.3  & yes & 3.28\,/\,9.11  & 0.16\,/\,0.47 \\
Hip muscle (Fig~\ref{fig:hip-muscle})
                    & 6,793  & 34,118  & 82  & 21 & 28.3  & yes & 7.41\,/\,21.27 & 0.39\,/\,1.85 \\
Hip bone            & 6,796  & 28,740  & 499 & 34 & 49.2  & no  & 4.12\,/\,14.27 & 0.25\,/\,1.30 \\
Liver               & 11,392 & 43,221  & 424 & 80 & 128.3 & no  & 9.00\,/\,33.68 & 0.21\,/\,4.81 \\
Hand surface (Fig~\ref{fig:spikyArap})   
                    & 11,829 & 49,751  & 456 & 31 & 43.8  & no  & 4.87\,/\,16.78 & 0.07\,/\,0.86
\end{tabular}
\end{center}
\end{table*}


\subsection{Hand muscles}

In our muscle hand example, we extracted 17 hand muscles 
from an MRI scan (Figure~\ref{fig:musclesInMRI}). 
We obtained the scan and the already extracted
bone meshes from~\cite{Wang:2019:HMA}; scan resolution is 0.5mm x 0.5mm x 0.5mm .
We considered two ``templates'', the first one from the
Centre for Anatomy and Human Identification at the University
of Dundee, Scotland~\shortcite{Dundee}, 
and the second one from Zygote~\cite{Zygote}.
We used the first one (Figure~\ref{fig:musclesInMRI}, left)
because we found it to 
be more medically accurate (muscles insert to correct bones).
Muscle anatomy of a human hand is challenging 
(Figure~\ref{fig:musclesInMRI}). 
We model all  muscle groups of the hand, namely the
thenar eminence (thumb),
hypothenar eminence (below little finger), 
interossei muscles (palmar and dorsal) (between
metacarpal bones), adductor pollicis 
(soft tissue next to the thumb, actuating thumb motion),
and lumbricals (on the side of the fingers at the fingers base). Our template models the correct number and general location of the
muscles, but there are large muscle shape differences between the template subject and the scanned subject (Figure~\ref{fig:teaser}).
We solve the optimization problem of 
Equations~\ref{eq:objectiveFunction}
and~\ref{eq:constraints} separately for each muscle, 
starting from the template mesh as the initial guess.
In our results, this produces muscles that match the  attachments and 
medical image constraints markers 
at 0.5 mm or better, which is at, or better than, the 
accuracy of the MRI scanner. 

\subsubsection{Marking the muscles in MRI scans}

During pre-processing, we manually mark as many 
reliable points as possible on the boundary
of each muscle ($\sim\!10-20$ landmarks and 
$\sim\!50-100$ ICP markers per muscle) 
in the MRI scans. 
 This process took approximately 5 minutes per
muscle.

\subsubsection{Attachments to bones}

The template muscles are modeled as triangle meshes. We build a tetrahedral mesh for each muscle.
Our tet meshes conform to the muscle's surface triangle mesh;
this requirement could be relaxed. For each muscle in the template, we attach its tet mesh
to the bones using soft constraints.
We do this by marking where on one or multiple bones this muscle inserts; to do so, we consulted
a medical doctor with specific expertise in anatomy. For each bone triangle mesh vertex that
participates in the insertion, 
we determine the material coordinates
(i.e., tet barycentric coordinates)
in the closest muscle tet.
We then form a soft constraint whereby this muscle material point
is linked to the  bone vertex position using a spring.

\subsubsection{Direct attempt using segmentation:}
We note that we have also attempted to model the muscle shapes directly using segmentation,
simply from an MRI scan. Recent work
has demonstrated that this can be done for hand bones~\cite{Wang:2019:HMA}, and we attempted
a similar segmentation approach  for muscles. However, given that the muscles touch each other
in many
places (unlike bones), the contrast in the MRI scan was simply not sufficient to discern the
individual muscles. Our conclusion is that a segmentation approach is not feasible for hand muscles, and one must use a pre-existing anatomically accurate template as in our work.

\subsubsection{Removing inter-penetrations of muscles}
\label{sec:collisions}

\begin{figure}[!t]
\centering
\includegraphics[width=0.8\hsize]{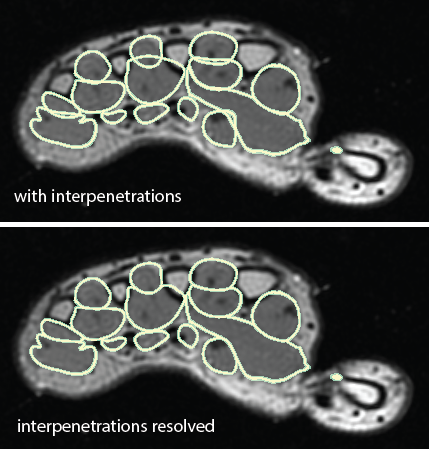}
\figspace{}
\vspace{-0.3cm}
\caption{ {\bf Interpenetration removal.} The yellow
lines are muscle cross-sections in this representative MRI slice of the hand.
It can be seen that our interpenetration removal method
successfully removes penetrations, without modifying the 
interpenetration-free sections of the muscles' boundary.
}
\vspace{-0.25cm}
\label{fig:removeInterpenetrations}
\end{figure}

Many hand muscles are in close proximity to one another and
several are in continuous contact. One strategy to resolve
contact would be to incorporate contact into our optimization
(Equations~\ref{eq:objectiveFunction} and~\ref{eq:constraints}). 
This approach is not very
practical, as unilateral constraints are very
difficult to optimize. Furthermore, such an approach
couples all (or most) muscles, and requires one to solve
an optimization problem with a much larger number of
degrees of freedom. It is extremely slow
at our resolution; it overpowers our machine.
Instead, we optimize each muscle
separately. Of course, the different muscles inter-penetrate each other,
which we resolve as follows.
For each muscle, we already positioned the MRI constraints
so that they are at the muscle boundary. Therefore, observe
that if our marker positioning and the solution to the optimization problem
were perfect, inter-penetration would be minimal or non-existent.
The marker placement is relatively straightforward at the boundary between a muscle
and another tissue (bone, fat, etc.) due to good contrast.
However, placing markers at the boundary between two continuously
colliding muscles is less precise, due to a lower MRI
contrast between adjacent muscles. This is the main cause
of the inter-penetrations. We remove the inter-penetrations 
with FEM simulation because it produces smooth organic
shape changes; note that alternatively, geometric approaches 
could also be used~\cite{Schmid:2009:MSM}.
Specifically in our work, for a pair of inter-penetrating muscles, we run
collision detection to determine the set of triangles of each
muscle that are inside the volume of the other muscle.
On each muscle, we then determine the set of tetrahedra that
are adjacent to the collision area. We then slightly enlarge this set,
by including any tet within a 5-ring of tets. We then run a FEM contact simulation 
just on these two tetrahedral sets on the two muscles. The FEM simulation
pushes the muscle surface boundaries apart, without displacing the rest
of the muscle (Figure~\ref{fig:removeInterpenetrations}). 
We handle contact islands of multiple muscles by running the above procedure
on two muscles, then for a third muscle against the first two muscles,
then the fourth against the first three, and so on. 



\subsection{Hip bone}
\label{sec:hipBone}

In our second anatomical example (Figure~\ref{fig:hipBone}), we apply our
method to a CT scan of the human right hip bone (pelvis).
We obtained the template from the human anatomy model
of Ziva Dynamics~\cite{ZivaAnatomy}, and the CT scan from
the ``KidneyFullBody'' medical image repository~\cite{Stephcavs}.
The template and the scanned hip bone differ substantially
in shape, and this is successfully captured by our method.

\subsection{Liver}
\label{sec:liver}

In our third anatomical example, we apply our
method to a CT scan of the human liver (Figure~\ref{fig:liver}).
We purchased a textured liver triangle mesh on 
TurboSquid~\cite{TurboSquid}. We subdivided it and 
created a tet mesh using TetGen~\cite{Hang:2011:TET}. This serves as our ``template''.
We used a liver CT scan from
the ``CHAOS'' medical image repository~\cite{CHAOSdata2019}.
We then executed our method to reshape the template tet mesh to match the CT scan.
Much like with the hip bone, our method successfully
models the large differences between the template and the 
scanned liver.
Finally, we embedded the TurboSquid triangle mesh into the template tet mesh,
and transformed it with the shape deformation of the tet mesh.
This produced a textured liver mesh (Figure~\ref{fig:liver}) that matches the CT scan.

\subsection{Hip muscle}
\label{sec:hip-muscle}

In our fourth anatomical example, we apply our
method to a MRI scan of a female human hip muscle (gluteus medius) (Figure~\ref{fig:hip-muscle}).
We obtained the data from The Cancer Imaging Archive (TCIA)~\cite{Clark:2013:TCIA}.
The image resolution is $384 \times 384 \times 240$ with voxel spacing of 1mm,
which is 2x coarser to the hand MRI dataset.
We use the template mesh from the human anatomy model of Ziva Dynamics~\cite{ZivaAnatomy}. 
We subdivided it and created a tet mesh for it using 
TetGen~\cite{Hang:2011:TET}. 
Because the muscle is attached to the hip bone and the leg bone, 
we needed to first extract the bones from the MRI scan; 
we followed the method described in~\cite{Wang:2019:HMA}.
Note that the subject in the Ziva Dynamics template is male.
The template and the scanned hip muscle differ substantially in shape, 
and this is successfully captured by our method.

\begin{figure}[!ht]
\centering
\includegraphics[width=1.0\hsize]{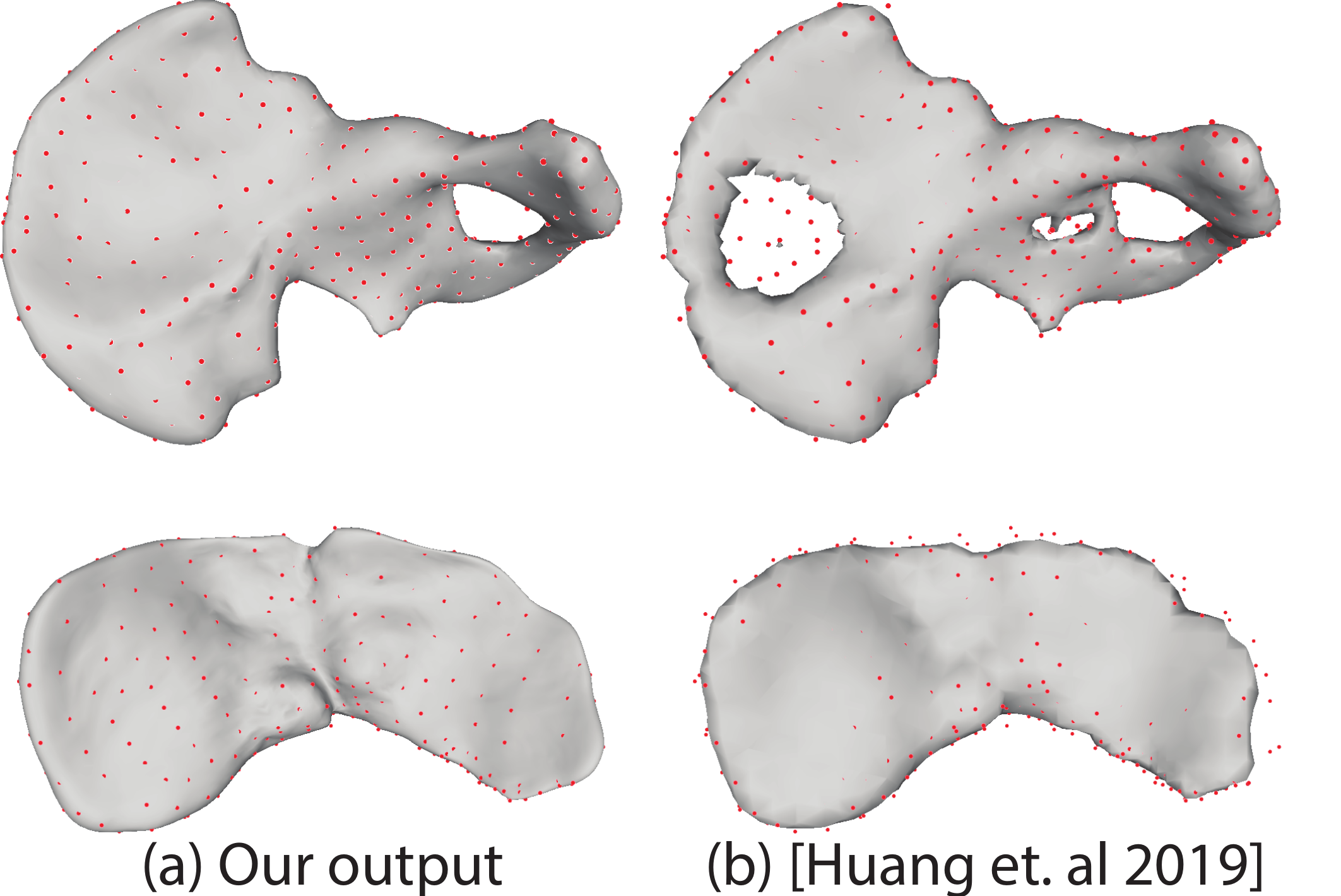}
\figspace{}
\vspace{-0.7cm}
\caption{ {\bf Comparison to~\cite{Huang:2019:VIP}.} 
Top row: hip bone. Bottom row: liver.
Red points are the markers from the CT scan.
We used the publicly available implementation of~\cite{Huang:2019:VIP} to compute
the normals, followed by screened Poisson surface reconstruction using the points and computed normals~\cite{Kazhdan:2013:SPS}.
We used this combination because it produced better results than running~\cite{Huang:2019:VIP} directly.
It can be seen that our method produces shapes that match the ground truth data more closely.
}
\vspace{-0.25cm}
\label{fig:vipss}
\end{figure}

\begin{figure}[!ht]
\centering
\includegraphics[width=1.0\hsize]{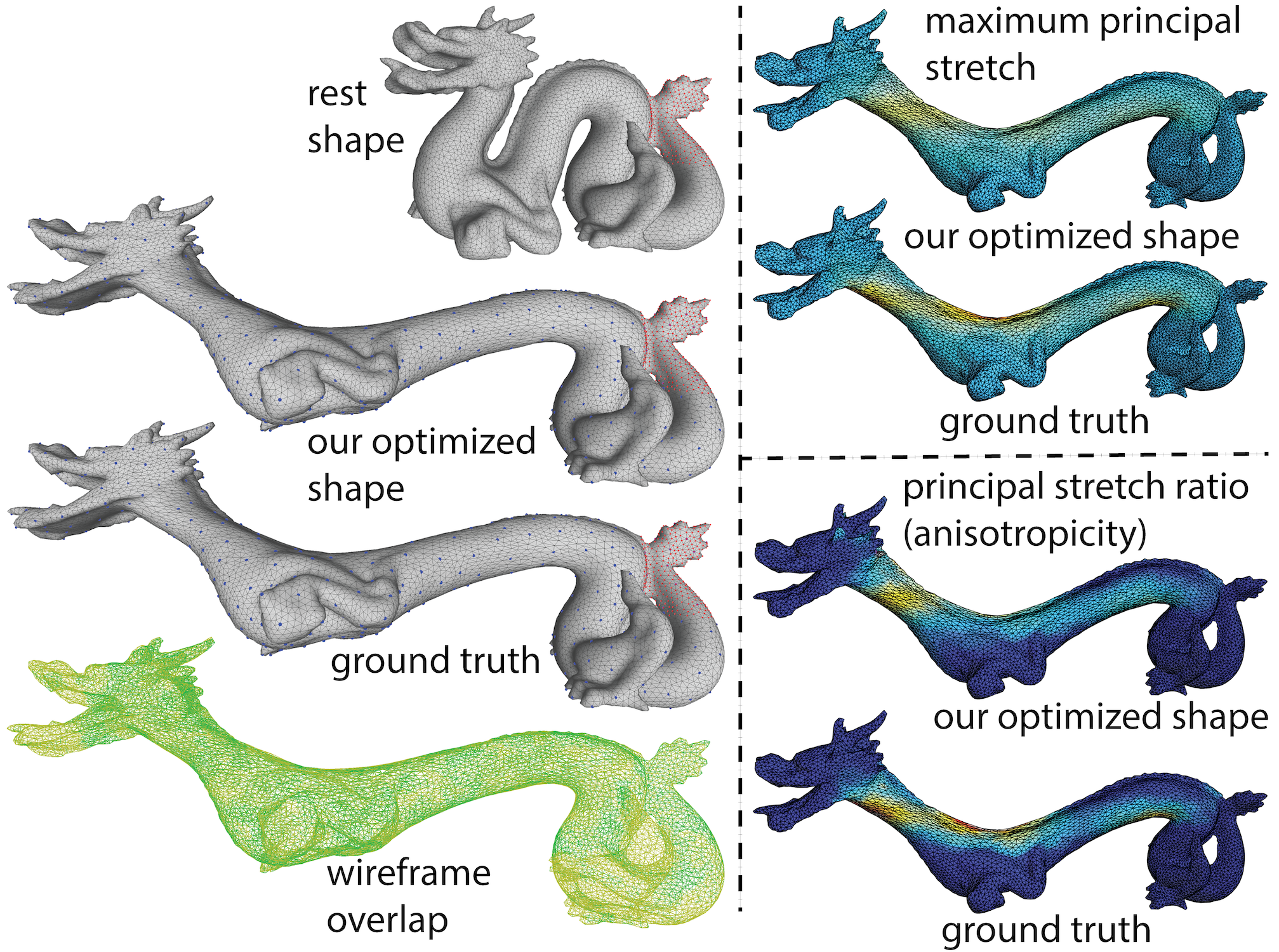}
\figspace{}
\vspace{-0.7cm}
\caption{ {\bf Quantitative evaluation on ground truth.} 
We first performed a dynamic FEM simulation with plasticity~\cite{Irving:2004:IFE},
whereby the back of the dragon is fixed (red), and the head was pulled to the
left with uniform force. This produced our ground truth plastic deformation gradients. 
We then selected 488 sparse triangle mesh vertices as landmarks and ICP markers, and
ran our method to compute $\BF{F_p}$ and shape $\BF{x}.$ It can be seen that $F_p$ and $x$
match the ground truth closely.
}
\vspace{-0.25cm}
\label{fig:plasticGroundTruth}
\end{figure}

\begin{figure}[!ht]
\centering
\includegraphics[width=1.0\hsize]{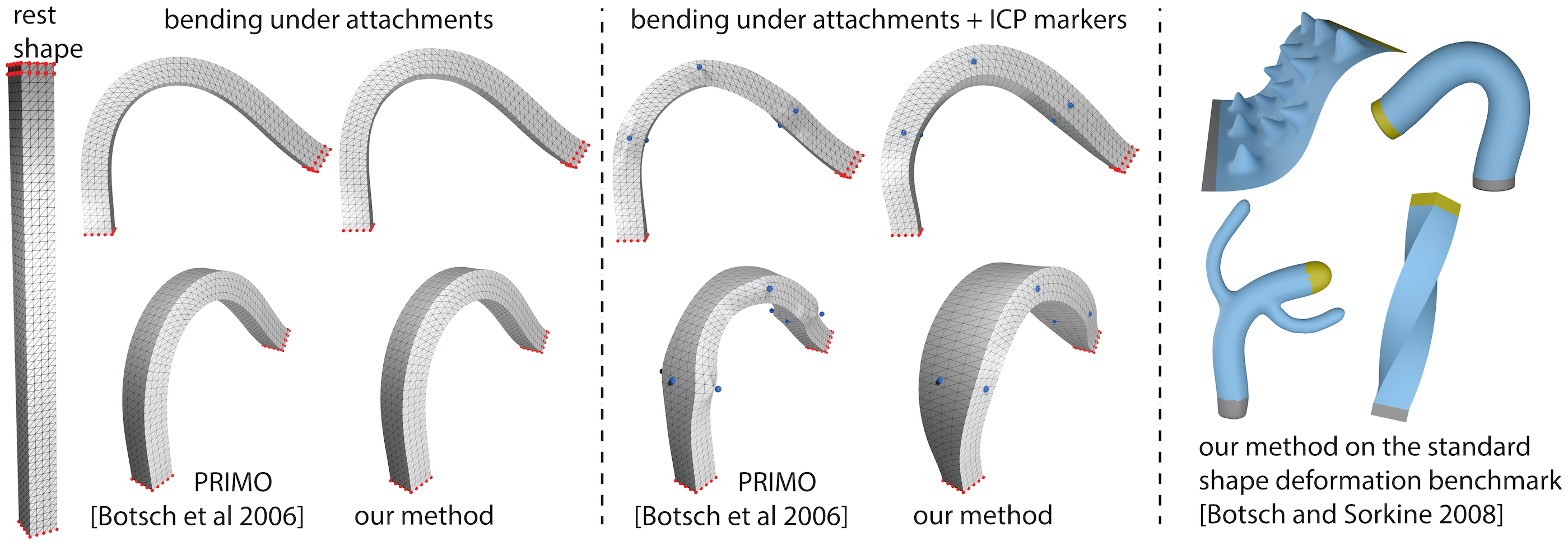}
\figspace{}
\vspace{-0.7cm}
\caption{ {\bf Comparison to surface-based methods.} 
We compare to the ``PRIMO'' method~\cite{Botsch:2006:PCP}
because this method is a good representative choice. It also 
has fewest artifacts in Figure 10 of the shape deformation survey~\cite{Botsch:2008:OLV}.
Our methods produces a clearly superior result in the
challenging scenario where the ICP markers were placed to anisotropically
stretch the beam in one transverse direction.
Our method also passes the standard shape deformation benchmark~\cite{Botsch:2008:OLV}.
}
\vspace{-0.25cm}
\label{fig:surfaceMethods}
\end{figure}

\subsection{Comparison to variational methods}
\label{sec:variational}

\begin{figure}[!ht]
\centering
\includegraphics[width=1.0\hsize]{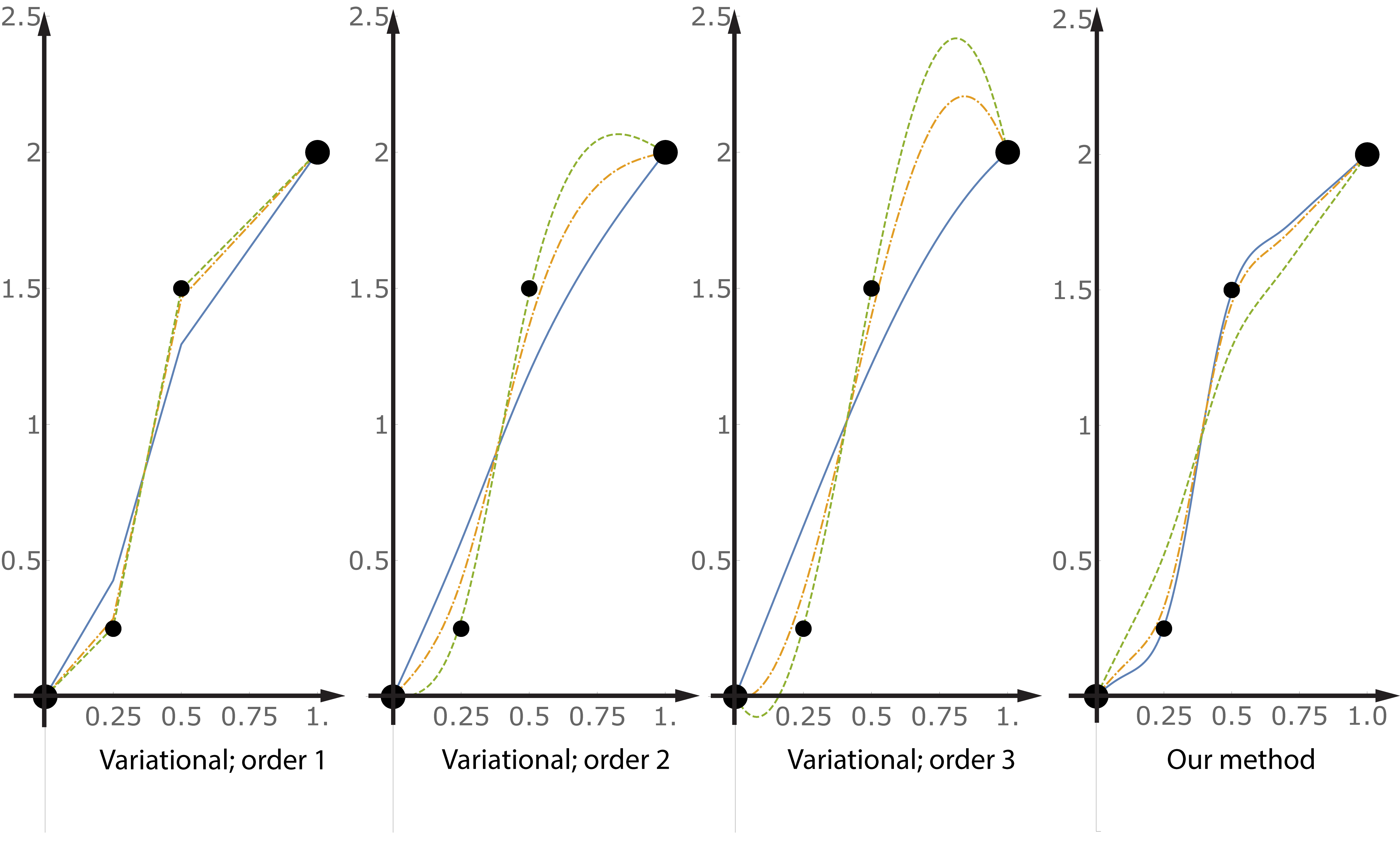}
\figspace{}
\vspace{-0.7cm}
\caption{ {\bf Comparison to variational methods.} 
A 1D string of length 1 is attached on both ends. 
The left attachment is fixed. The right attachment is moved to coordinate 2.
The spring thus stretches longitudinally 
while trying to obey the two landmarks at $t=1/4$ and $t=1/2.$
Y-axis gives the deformed 1D position of the corresponding material point on the X-axis.
Big and small dots denote the attachments and landmarks, respectively.
For each method, we plot the result under a few representative parameters.
With variational methods, one has to either give up meeting the landmarks,
or the curve becomes wiggly. Similar behavior can also be observed in 
3D variational results (Figure~\ref{fig:kobbelt}). Our method produces a deformation profile
whereby the slope (total 1D deformation gradient) on all three subintervals
$[0,1/4],[1/4,1/2],[1/2,1]$ approximately matches the slope implied by the attachments
and landmarks (1, 5, 1, respectively).
Note that the shown output curves are only $\mathcal{C}^{r-1}$ and
not in $\mathcal{C}^r$ at the two juncture points
$1/4,1/2;$ however, their $r$-th derivative is integrable and they
are the optimal Cauchy limit of a score-decreasing sequence of curves in $\mathcal{C}^r.$
}
\vspace{-0.25cm}
\label{fig:variational}
\end{figure}

We compare our method to variational shape modeling methods on an illustrative 1D example. 
Note that Figure~\ref{fig:kobbelt} gave a comparison on 3D muscle geometry.
Consider an elastic 1D line segment whose neutral shape
is the interval $[0,1],$ and study its longitudinal
1D deformation under the following setup.  
Let us prescribe hard attachments
whereby we attach endpoint $0$ to position $0,$
and endpoint $1$ to position $2.$ Furthermore, assume
landmarks whereby point $1/4$ is at $1/4,$
and point $1/2$ is at $3/2.$ Effectively in this setup,
we are specifying that the subintervals $[0,1/4]$ and
$[1/2,1]$ do not stretch, whereas the subinterval $[1/4,1/2]$
stretches from its original length of $1/4$ to $5/4,$ (5x stretch).
A variational formulation of order $r$ is,
\begin{gather}
\min_{x(t)\in\mathcal{C}^r} \int_0^1 \bigl(\frac{d^r\!x}{dt}\bigr)^2 dt + 
\alpha \Bigl(\bigl(x(\frac{1}{4})-\frac{1}{4}\bigr)^2 + \bigl(x(\frac{1}{2}) - \frac{3}{2}\bigr)^2\Bigr),
\end{gather}
where $\mathcal{C}^r$ denotes all functions $[0,1]\to \RR$ whose derivatives exist and are continuous
up to order $r.$
We solved these problems analytically in Mathematica
for $r=1,2,3,$ each time for three representative values $\alpha,$
and compared them (see Figure~\ref{fig:variational}) to our method in 1D,
\begin{gather}
\nonumber
\min_{f_p,\, x(t)\in\mathcal{C}^r} \int_0^1 \ddot{f_p}^2 dt + \alpha 
\Bigl(\bigl(x(\frac{1}{4})-\frac{1}{4}\bigr)^2 + \bigl(x(\frac{1}{2}) - \frac{3}{2}\bigr)^2\Bigr)\\
+ \beta \int_0^1 \bigl(\frac{\dot{x}}{f_p} - 1 \bigr)^2 dt\\
\textrm{s.\ t.\ } x(t)=\argmin_{x(t)\in\mathcal{C}^r} \int_0^1 \bigl(\frac{\dot{x}}{f_p} - 1 \bigr)^2 dt. 
\end{gather}
Observe that, in the same vein as in 3D, we can decompose $\dot{x} = f_e f_p,$
whereby scalar functions $f_e$ and $f_p$ and the $\beta$ term
are the equivalents of the elastic and plastic
deformation gradients, and the elastic energy, respectively.
It can be seen that our method produces a better fit to the data
and a significantly less wiggly solution, compared to variational methods
(Figure~\ref{fig:variational}).

\section{Conclusion}

We gave a shape deformation method that can model objects
undergoing large spatially varying strains. Our method works
by computing a plastic deformation gradient at each tet, such
that the mesh deformed by these plastic deformation gradients
matches the provided sparse landmarks and closest-point constraints.
We support both constrained and unconstrained objects; the latter are supported
by giving a numerical method to solve large sparse linear systems of equations
with a known nullspace. 
We applied our method to the extraction of shapes of organs from medical images.
Our method has been designed to extract as much information
as possible from MRI images, despite the  soft boundaries
between the different  organs. We extracted hand muscles,
a liver, a hip bone and a hip muscle.

Our method does not require a dense target mesh; only a sparse set of observations
is needed. If a dense target mesh is available, the problem becomes somewhat
easier, as one can then use standard ICP algorithms. However, medical images contain many ambiguities 
and regions where there is not a sufficient contrast to clearly disambiguate two adjacent medical organs;
making it impractical to extract dense target meshes.
We apply our method to solid objects, but our plastic strain shape deformation 
method could also be used for shells (cloth). Doing so would require formulating the elastic energy of a plastically deformed FEM cloth, 
and computing the energy gradients and Hessians with respect to the plastic parameters. 
The size of the small square dense matrix that we need to invert in our incremental 
solve is three times the number of markers.
While we easily employed up to a thousand of markers in our work, our method will slow
down under a large number of markers.
We do not re-mesh our tet meshes during the optimization. If the plastic strain causes
some tetrahedra to collapse or nearly collapse, this will introduce numerical instabilities.
Although not a problem in our examples (see Figure~\ref{fig:minDihedralAngles}), 
such situations could be handled by re-meshing
the tet mesh during the optimization~\cite{Bargteil:2007:FEM}.
Our method requires a plausible template with a non-degenerate tet mesh.
Re-meshing is important future work as it could extend the reach of our method, enabling one to start 
the optimization simply from a sphere tet mesh.

%
%
%
%

\bibliographystyle{ACM-Reference-Format}
\bibliography{hand-muscles}

\appendix

\section{Plastic Strain Laplacian and its nullspace}
\label{app:laplacian}

Let $\BF{L^\textrm{sc}}\in\RR^{m\times m}$ denote the  discrete 
mesh Laplacian for \emph{scalar} fields  on
mesh tetrahedra ~\cite{Zhang:2004:DCP}, 
\begin{equation}
\BF{L^\textrm{sc}}_{i, j} = 
\left\{
        \begin{array}{ll}
                \#\textrm{adjacent\ tets\ }  & \mbox{if } i=j, \\
                -1 & \mbox{if } i\ne j\textrm{,\ and\ } i,j\textrm{\ are\ adjacent},\\
                0 & \textrm{otherwise}.
        \end{array}
\right.
\label{eq:discreteL}
\end{equation}
Given a plastic strain state $\BF{s}
=[\BF{s_1},\BF{s_2},\BF{s_3},\BF{s_4},\BF{s_5},\BF{s_6}]\in\RR^{6m},$
where $\BF{s_i}\in\RR^m,$ define the \emph{plastic strain Laplacian} 
\begin{equation}
\BF{L}\BF{s}=[\BF{L^\textrm{sc}}\BF{s_1},\sqrt{2}\BF{L^\textrm{sc}}\BF{s_2},
\sqrt{2}\BF{L^\textrm{sc}}\BF{s_3},\BF{L^\textrm{sc}}\BF{s_4},
\sqrt{2}\BF{L^\textrm{sc}}\BF{s_5},\BF{L^\textrm{sc}}\BF{s_6}],
\end{equation}
where the $\sqrt{2}$ were added to account for the fact that $s_2,s_3,s_5$
control two entries in the symmetric matrix $F_p\in\RR^{3\times 3}.$

\noindent \textbf{Lemma:}
Assume that the tet mesh $\mathcal{M}$ has a single connected component.
Then, the nullspace of $\BF{L}$ is 6-dimensional
and consists of vectors $\psi_i=[\BF{s_1},\ldots,\BF{s_6}]$
where $\BF{s_j}\in\RR^m$ is all ones when $j=i,$ and all zeros otherwise.\\
\noindent\textbf{Proof:}
First, observe that $\BF{L^\textrm{sc}}$ is symmetric positive semi-definite
with a single nullspace vector, namely the vector of all 1s.
This follows from the identity
\begin{equation}
\BF{x}^T \BF{L^\textrm{sc}} \BF{x}=
\sum_{i\textrm{\ and\ }j\textrm{\ adjacent}}(x_i-x_j)^2,
\end{equation}
i.e., $\BF{x}^T \BF{L^\textrm{sc}} \BF{x}= 0$ is only possible if all $x_i$ are the same.

We have $0=\BF{s}^T\BF{L}\BF{s}=
\sum_{i=1}^6 \xi_i \BF{s_i}^T\BF{L^\textrm{sc}} \BF{s_i},$
where $\xi_i=\sqrt{2}$ for $i=2,3,5;$ and $1$ otherwise.
Because $\BF{L^\textrm{sc}}$ is symmetric positive semi-definite,
each $\BF{s_i}$ must either be $\BF{0}$ or a non-zero
nullspace vector of $\BF{L^\textrm{sc}},$ i.e., a vector
of all 1s. A linearly independent orthonormal nullspace
basis emerges when we have a vector for all 1s for exactly
one $i.$ There are 6 such choices, giving the vectors $\psi_i;$
we normalize them by dividing with $\sqrt{m}.$ 
$\hfill\blacksquare$

\section{First and Second Derivatives of elastic 
energy with respect to plastic strain}
\label{app:appendixDerivatives}

For convenience, 
we denote $F_{p,i}$ as $i-$th entry of the 
vector $\text{vec}(F_p) \in \mathcal{R}^9$.
The first-order derivatives are
\begin{gather}
\PP{\mathscr{E}}{x_i}=
V\PP{\psi}{F_e}\colon \PP{F_e}{x_i} = V P \colon \PP{F_e}{x_i},\\
\PP{\mathscr{E}}{F_{p,i}}=
V \PP{\psi}{F_{p,i}}+\PP{V}{F_{p_i}}\psi=
V P \colon \PP{F_e}{F_{p,i}}+\PP{V}{F_{p_i}}\psi,\quad\textrm{where}\\
\PP{F_e}{x_i}=\PP{F}{x_i} F_p^{-1},\quad\PP{F_e}{F_{p,i}}=
F \PP{F_p^{-1}}{F_{p,i}},\quad\textrm{and}\\
\PP{V}{F_{p,i}}=\PP{\left|F_p\right|}{F_{p_i}} V_0.
\end{gather}
Here, $P$ is the first Piola-Kirchhoff stress tensor and $\PPI{F}{x}$ is a constant matrix commonly used in the equations for FEM simulation.
For the second-order derivatives, we first compute $\PPPI{\mathscr{E}}{x}$. 
This is the tangent stiffness matrix in the FEM simulation under a fixed $F_p$.
It is computed as
\begin{equation}
\PPPP{\mathscr{E}}{x_i}{x_j}=V \PP{F_e}{x_j}^T \colon \PP{P}{F_e} \colon \PP{F_e}{x_i}.
\end{equation}
Here, $\PPI{P}{F_e}$ is a standard term in FEM nonlinear elastic simulation; it only depends on the strain-stress law (the material model).
Next, we compute $\PPPPI{\mathscr{E}}{x}{F_p},$
\begin{gather}
\PPPP{\mathscr{E}}{x_j}{F_{p,i}}=
\PP{V}{F_{p,i}} \left(P \colon \PP{F_e}{x_i}\right)+\\
V\PP{F_e}{F_{p,j}}^T\colon\PP{P}{F_e}\colon\PP{F_e}{x_i}+
V P \PPPP{F_e}{x_i}{F_{p,j}},\quad\textrm{where}\\
\PPPP{F_e}{x_i}{F_{p,j}}=\PP{F}{x_i}\PP{F_p^{-1}}{F_{p,j}}.
\end{gather}
Finally, we have 
\begin{gather}
\PPPP{\mathscr{E}}{F_{p,i}}{F_{p,j}}=
\PPPP{V}{F_{p,i}}{F_{p,j}}\psi+\\
V \left(P \colon \PPPP{F_e}{F_{p,i}}{F_{p,j}}+
\PP{F_e}{F_{p,j}}^T \colon \PP{P}{F_e} \colon \PP{F_e}{F_{p,i}}\right)+\\
\PP{V}{F_{p,j}}\PP{\psi}{F_{p,i}}+\PP{V}{F_{p,i}}\PP{\psi}{F_{p,j}},\quad\textrm{where}\\
\PPPP{V}{F_{p,i}}{F_{p,j}} = \PPPP{\left|F_{p}\right|}{F_{p,i}}{F_{p,j}} V_0,\\
\PPPP{F_e}{F_{p,i}}{F_{p,j}} = F \PPPP{F_p^{-1}}{F_{p,i}}{F_{p,j}}.
\end{gather}
The quantities $\psi, P$ and $\PPI{P}{F_e}$ are determined 
by the chosen elastic material model. After computing the above derivatives, 
there is still a missing link between $F_p$ and $s$. 
Because we want to directly optimize $s$, we also need the derivatives 
of $\mathscr{E}(F_p(s),x)$ with respect to $s$. 
From Equation~\ref{eq:Fp-s} we can see that $F_p$ is 
linearly dependent on $s.$ Therefore, so we can define
a matrix $Y$ such that $\text{vec}(F_p)=Y s$. 
Then all the derivatives can be easily transferred to 
derivation by $s$ by multiplying with $Y$.

\section{Proof of Singular Lemma}
\label{app:singularLemma}

Statement (i) follows from well-known linear algebra facts
$\mathcal{R}(A)=\mathcal{N}(A^T)^\perp$ and 
$\mathrm{dim}(\mathcal{N}(A))+\mathrm{dim}(\mathcal{R}(A))=p,$
and the symmetry of $A.$ As per (ii), 
$A$ maps $\mathcal{R}(A)$ into itself, and no vector from 
$\mathcal{R}(A)$ maps to zero, hence
the restriction of $A$ to $\mathcal{R}(A)$ is invertible, establishing 
a unique solution to $Ax=b$ with the 
property that $x\perp \psi_i$ for all $i=1,\ldots,k.$
This unique solution is the minimizer of 
\begin{gather}
\min_{x} \frac{1}{2} x^T A x - b^T x\\
\textrm{s.t.}\quad \psi_i^T x =0\ \textrm{for\ all\ } i=1,\ldots,k.
\end{gather}
When expressed using Lagrange multipliers, 
this gives Equation~\ref{eq:singularSolve}.
Suppose $x=n+r$ is another solution and $n\in\mathcal{N}(A)$
and $r\in\mathcal{R}(A)$. Then $b=Ax=Ar$ and hence 
$r$ is the unique solution from Equation~\ref{eq:singularSolve}.
The vector $n$ can be an arbitrary nullspace vector, 
proving the last statement of (ii). 
As per (iii), suppose we have $0 = B(n+r) = 
Ar + \sum_{i=1}^k \frac{\lambda_i}{\alpha_i} (\psi_i^T n)\psi_i .$
Observe that the first summand is in $\mathcal{R}(A)$ and the second
in $\mathcal{N}(A).$ Hence, $B(n+r)$
can only be zero if both summands are zero. $Ar=0$ implies
$r=0.$ The second summand can only be zero if $n\perp \psi_i$ for each
$i,$ which implies that $n=0.$ Hence, $B$ is invertible.
The last statement of (iii) can be verified by expanding
$\bigl(A+\sum_{i=1}^k \alpha_i \psi_i \psi_i^T\bigr)
\bigl(x+\sum_{i=1}^k \frac{\lambda_i}{\alpha_i} \psi_i\bigr).
\hfill\blacksquare$

\section{Proof of Nullspace Lemma}
\label{app:nullspaceLemma}

\setlength{\columnsep}{9pt}
\begin{wrapfigure}[19]{r}{0.25\textwidth}
{\vspace{-0.4cm}}
\includegraphics[width=0.25\textwidth]{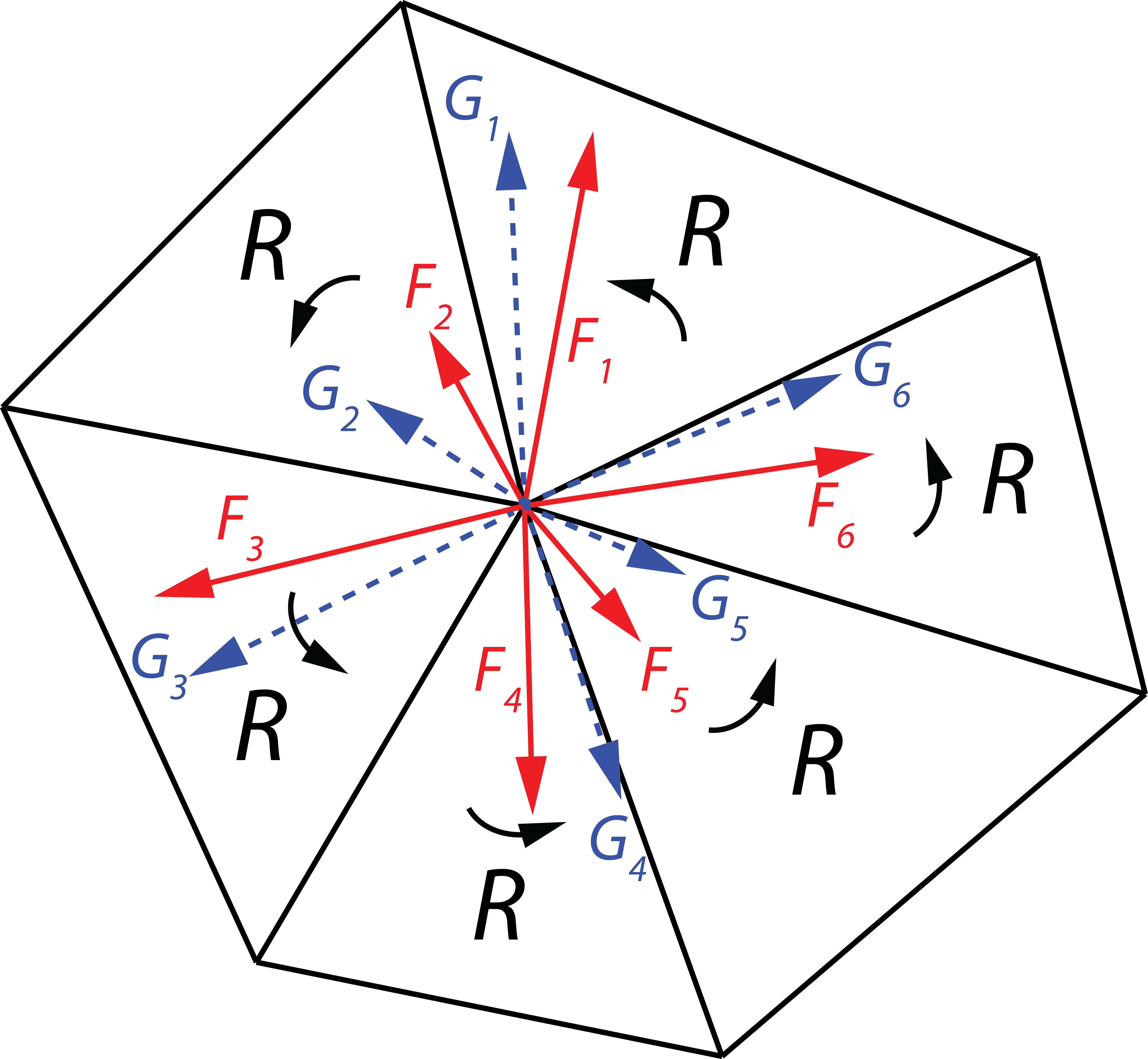}
\figspace{}
{\vspace{-0.6cm}}
\caption{
\textbf{Illustration of the nullspace proof.}
The original forces $F_i$ sum to zero.
We have $G_i = R F_i;$ note that $R$ is the same for all tets.
Therefore, the rotated forces $G_i$ also sum to zero.
Hence there is no change in the internal elastic force under a rotation,
i.e., infinitesimal rotations are in the nullspace of $\BF{K}.$
}
\label{fig:singularK}
\end{wrapfigure}
We are trying to prove that
$\BF{K}(\BF{F_p}(\BF{s}),\BF{x})$
is 6-dimensional for any $\BF{x}$ that solves
$\BF{f}_{\textrm{e}}(\BF{s},\BF{x})= 0.$ 
First, if $\BF{x}$ is a solution,
then translating all vertices
of the object by the same constant 3-dimensional vector is also
a solution. This means that vector $\psi_i := (e_i,e_i,\ldots,e_i)$
is in the nullspace of $\BF{K},$ where $e_i\in\RR^3$ is the $i$-th
standard basis vector, for $i=1,2,3.$
Now, suppose we rotate the object with an infinitesimal rotation
$X\mapsto X + e_i\times X.$ 
Observe that for general plastic strains $\BF{s},$
the elastic forces in each individual tet are not zero
even in the equilibrium $\BF{x};$ but the contributions
of elastic forces on a tet mesh vertex from all adjacent
tets sum to zero. As we rotate the object,
the forces contributed by adjacent tets to a specific tet mesh vertex
rotate by the same rotation in each tet. 
Therefore, as these forces sum to zero, they continue to sum to zero
even under the rotation (see Figure~\ref{fig:singularK}).
This means that the
vector of infinitesimal displacements 
$\psi_{3 + i} := [e_i\times x_1, e_i\times x_2, \ldots, e_i\times x_n]$ 
induced by the infinitesimal rotation is in the nullspace of $\BF{K},$ for each $i=1,2,3.$
Here, $x_i$ are the components
of $\BF{x}=[x_1, x_2, \ldots, x_n].$ The vectors $\psi_i,\ i=1,2,3,4,5,6,$ form
the nullspace of $\BF{K}.$ $\hfill\blacksquare$

Finally, we inform the reader that the nullspace of 
$\BF{K}(\BF{F_p}(\BF{s}),\BF{x})$ is only 3-dimensional if $\BF{x}$ is
\emph{not} an elastic equilibrium. In this case, only translations
are in the nullspace. Infinitesimal rotations are not in the nullspace
because under an infinitesimal rotation, the non-zero elastic
forces $\BF{f_\textrm{e}}$ rotate, i.e., they do not remain the same.
The assumption of $\BF{x}$ being the equilibrium shape is therefore
crucial (and is satisfied in our method).

\section{Second derivative of polar decomposition}
\label{app:secondPolar}

To compute the second-order derivatives, we differentiate
\begin{gather}
\PP{F}{F_i}=\PP{R}{F_i}S+R\PP{S}{F_i},\\
\PPPP{F}{F_i}{F_j}=\PPPP{R}{F_i}{F_j}S+ 
\PP{R}{F_i}\PP{S}{F_j}+\PP{R}{F_j}\PP{S}{F_i}+R\PPPP{S}{F_i}{F_j},\\
\PPPP{R}{F_i}{F_j}=\left(-R\PPPP{S}{F_i}{F_j}-\PP{R}{F_j}\PP{S}{F_i}-\PP{R}{F_i}\PP{S}{F_j}\right)S^{-1}.
\end{gather}
To compute $\PPPPI{R}{F_i}{F_j}$, we need to compute $\PPPPI{S}{F_i}{F_j}$ first.
This can be derived in the same way as for $\PPI{S}{F_i}$.
Starting from Equation~\ref{eq:S-1st-deriv}, we have
\begin{equation}
\PPPP{F^TF}{F_i}{F_j}=\PPPP{S}{F_i}{F_j}S+\PP{S}{F_i}\PP{S}{F_j}+\PP{S}{F_j}\PP{S}{F_i}+S\PPPP{S}{F_i}{F_j}.
\end{equation}
We can now solve a similar Sylvester equation
\begin{gather}
\text{vec}(\PPPP{S}{F_i}{F_j})=\left(S \oplus S\right)^{-1}\text{vec}(C),\\
C=\PPPP{F^TF}{F_i}{F_j}-\PP{S}{F_i}\PP{S}{F_j}-\PP{S}{F_j}\PP{S}{F_i}.
\end{gather}

\section{Formulas for $\BF{A_k}$,$\BF{b_k}$,$\BF{c_k}$ (Equation~\ref{eq:energy-full})}
\label{app:Akbkck}

Let the landmark $k$ be embedded into a tetrahedron $t_k$ with  
barycentric weights $\left[w^{k}_1, w^{k}_2, w^{k}_3, w^{k}_4\right]$. We have
\begin{align}
\BF{A_k}&= \begin{bmatrix}
w^{k}_1 I_3 & w^{k}_2 I_3 &  w^{k}_3 I_3 &  w^{k}_4 I_3
\end{bmatrix} S^{k} \\
\BF{b_k}&=-y_k,
\end{align}
where $y_k$ is the landmark's target position,
$I_3 \in \RR^{3 \times 3}$ is the identity matrix, and
$S^{k} \in \RR^{12\times 3n}$ is
a selection matrix that selects the positions of 
vertices of $t_k$.
The scalar $\BF{c_k}$ is the weight of the landmark $k.$
An equivalent formula applies to ICP markers and attachments. 

\end{document}